# Design and Analysis of the REESSE1+ Public Key Cryptosystem v2.21 *

Shenghui Su [1] and Shuwang Lü [2]

[1] College of Computers, Beijing University of Technology, Beijing 100124, P.R.China

[2] Graduate School, Chinese Academy of Sciences, Beijing 100039, P.R.China

**Abstract**: In this paper, the authors give the definitions of a coprime sequence and a lever function, and describe the five algorithms and six characteristics of a prototypal public key cryptosystem which is used for encryption and signature, and based on three new problems and one existent problem: the multivariate permutation problem (MPP), the anomalous subset product problem (ASPP), the transcendental logarithm problem (TLP), and the polynomial root finding problem (PRFP). Prove by reduction that MPP, ASPP, and TLP are computationally at least equivalent to the discrete logarithm problem (DLP) in the same prime field, and meanwhile find some evidence which inclines people to believe that the new problems are harder than DLP each, namely unsolvable in DLP subexponential time. Demonstrate the correctness of the decryption and the verification, deduce the probability of a plaintext solution being nonunique is nearly zero, and analyze the exact securities of the cryptosystem against recovering a plaintext from a ciphertext, extracting a private key from a public key or a signature, and forging a signature through known signatures, public keys, and messages on the assumption that IFP, DLP, and LSSP can be solved. Studies manifest that the running times of effectual attack tasks are greater than or equal to $O(2^n)$ so far when $n$ = 80, 96, 112, or 128 with $\lg M \approx$ 696, 864, 1030, or 1216. As viewed from utility, it should be researched further how to decrease the length of a modulus and to increase the speed of the decryption.

## 1 Introduction

The trapdoor functions for RSA [1] and ElGamal [2] public key cryptosystems [3] are computationally one-way [4][5], which indicates that there always exists one sufficiently large setting of the security dominant parameter that makes utilization of a cryptosystem feasible and breaking of the cryptosystem infeasible in polynomial time [6]. Taking RSA based on the integer factorization problem (IFP) as an example, when the bit-length of a RSA modulus reaches 1024, attack is infeasible but encryption and decryption are feasible in polynomial time. Such a security is referred to as asymptotic security, which is distinguished from exact security or concrete security. The exact security is practice-oriented, and aims at giving more precise estimates of time complexities of attack tasks [7].

In some public key cryptosystems, trapdoor functions can prevent a related plaintext from being recovered from a ciphertext, but cannot prevent a related private key from being extracted from a public key. For instance, in the MH knapsack cryptosystem [8], the subset sum problem (SSP) which contains the trapdoor information cannot prevent a MH private key from being extracted from $c_i \equiv a_i W$ (% $M$) by the method of accumulating points of minima [9], and moreover, when a related knapsack density is less than 1, the low-density SSP (LSSP) will degenerate to a polynomial time problem from a NPC problem owing to the $L^3$ lattice base reduction algorithm [10] [11] which is employed for finding the shortest vector or an approximately shortest vector in a lattice [12].

Along with the elevation of a computer′s CPU speed, the security dominant parameter of a cryptosystem becomes larger and larger. For instance, the bit-length of modulus of the ElGamal cryptosystem based on the discrete logarithm problem (DLP) is already up to 1024. We think that there are currently four manners of decreasing the bit-lengths of security dominant parameters and meantime increasing the one-wayness of related trapdoor functions.

The first manner is to transplant known cryptosystems to a complex algebraic system from a simple one — the elliptic curve analogue of ElGamal referable to elliptic curve cryptography (ECC) for example. By now, any effectual algorithm which can find out generic elliptic curve discrete logarithms in time being subexponential in the bit-length of a modulus has not been discovered yet [13].

Theoretically, almost every existing cryptosystem may have an elliptic curve analogue. However, not every analogue can bring the same effect as the analogue of ElGamal — the analogue of RSA whose security still relies on the two large prime factors [14] for example.

The second manner is to design cryptosystems over polynomial rings — the NTRU cryptosystem for example. The shortest vector problem (SVP) is the security bedrock of NTRU since it is impossible to seek an NTRU secret polynomial or an NTRU plaintext polynomial through the $L^3$ lattice base reduction on condition that two special parameters $c_h$ and $c_m$ are fitly selected [15].

The third manner is to construct cryptosystems based on the tame automorphism of multivariate quadratic polynomials over a small field — the TTM scheme [16] and the TTS scheme [17] ordinarily referred to as the multivariate cryptosystems for example.

The fourth manner is to devise cryptosystems over small prime fields through discovering or constructing new one-way computational problems which should be harder than IFP, DLP, or LSSP in light of the found evidence. Some threads of the fourth manner are given in this paper.

* Manuscript first received 15 Nov 2006, and last revised 26 Dec 2012. In v2.21, the algorithms of REESSE1+ are intact, and only the computation of density of an ASSP Knapsack is corrected (Appendix B).

It occurs in *Theoretical Computer Science,* v426-427, Apr 2012, pp. 91-117.

This research is supported by MOST with Project 2007CB311100 and 2009AA01Z441. Corresponding email: reesse@126.com.

The paper has five novelties: ① Gives the definitions and properties of a coprime sequence, a lever function, and a bit shadow; ② offers three new computational problems — the multivariate permutation problem (MPP) $C_i \equiv (A_i W^{\ell(i)})^\delta$ (% $M$), the anomalous subset product problem (ASPP) $\bar{G} \equiv \prod_{i=1}^{n} C_i^{b_i}$ (% $M$), and the transcendental logarithm problem (TLP) $y \equiv x^x$ (% $M$), proves that the three problems are computationally at least equivalent to DLP in the same prime field each, and finds some evidence which inclines people to believe that the three problems are separately harder than DLP; ③ over a prime field, designs the five algorithms of a prototypal public key cryptosystem called REESSE1+; ④ analyzes the security of REESSE1+; ⑤ proposes and proves the double congruence theorem.

MPP which owns the indeterminacy assures the security of a private key and the signature algorithm. ASPP as a trapdoor function which can resist the $L^3$ lattice base reduction assures the security of a ciphertext. TLP protects a private key against being extracted from a signature, and moreover it concerts with a form of the polynomial root finding problem (PRFP) $ax^n + bx^{n-1} + cx + d \equiv 0$ (% $\bar{M}$) with $a \neq 0, 1$, $|b| + |c| \neq 0$, and $d \neq 0$ to assure the security of a signature. Provable security by reduction is appreciable, but not sufficient, and thus the exact security of REESSE1+ should be analyzed.

It is not difficult to understand that REESSE1+ is essentially a multiproblem cryptosystem. The security of a multiproblem cryptosystem is equivalent to the complexity of what is easiest solved in all the problems. Additionally, MPP contains the four variables almost independent, and therefore, in a broad sense, REESSE1+ may be regarded as multivariate. A multiproblem cryptosystem must be a multivariate cryptosystem because only multiple variables can bring multiple problems.

REESSE1+ is different from REESSE1 which has a ciphertext $\bar{G}_1 = \prod_{i=1}^{n} C_i^{b_i}$ % $M$ with $C_i = A_i W^{\ell(i)}$ % $M$ [18] and an insecure signature ($U = (UQH)^S$ % $M$, $V = V^T$ % $M$) [19], and also different from the Naccache-Stern cryptosystem which has a ciphertext $c = \prod_{i=1}^{n} v_i^{b_i}$ % $M$ with $v_i = p_i^{1/s}$ % $M$ [20]. It will be significant in untouched areas.

We know that in a quantum computational model, IFP and DLP are already solved in polynomial time [21], and naturally, whether MPP, ASPP, and TLP can be solved in polynomial time on a quantum computer is interesting. Besides, TLP as a primitive problem cannot be converted into a discrete logarithm problem, which indicates that one can design other signature schemes over a small prime field by using TLP or its variety $y \equiv (gx)^x$ (% $M$).

Throughout the paper, unless otherwise specified, $n \geq 80$ is the bit-length of a block or the item-length of a sequence, the sign % means 'modulo', $\bar{M}$ means '$M - 1$' with $M$ prime, lg$x$ denotes the logarithm of $x$ to the base 2, $\neg$ does the opposite value of a bit, $\mathcal{P}$ does the maximal prime allowed in coprime sequences, $|x|$ does the absolute value of a number $x$, $\|x\|$ does the order of an element $x$ % $M$ or the size of a set $x$, and gcd($a$, $b$) represents the greatest common divisor of two integers. Without ambiguity, '% $M$' is usually omitted in expressions.

# 2 A Coprime Sequence, a Lever Function, and a Bit Shadow

***Definition 1:*** If $A_1, \ldots, A_n$ are $n$ pairwise distinct positive integers such that $\forall\, A_i, A_j$ ($i \neq j$), either gcd($A_i$, $A_j$) = 1 or gcd($A_i$, $A_j$) = $F \neq 1$ with $(A_i / F) \nmid A_k$ and $(A_j / F) \nmid A_k$ $\forall\, k \neq i, j \in [1, n]$, these integers are called a coprime sequence, denoted by $\{A_1, \ldots, A_n\}$, and shortly $\{A_i\}$.

Notice that the elements of a coprime sequence are not necessarily pairwise coprime, but a sequence whose elements are pairwise coprime is a coprime sequence.

***Property 1:*** Let $\{A_1, \ldots, A_n\}$ be a coprime sequence. If randomly select $m \in [1, n]$ elements from $\{A_1, \ldots, A_n\}$, and construct a subsequence $\{A_{x_1}, \ldots, A_{x_m}\}$ also called a subset, then the subset product $G = \prod_{i=1}^{m} A_{x_i} = A_{x_1} \ldots A_{x_m}$ is uniquely determined, namely the mapping from $\{A_{x_1}, \ldots, A_{x_m}\}$ to $G$ is one-to-one.

*Proof:* By contradiction.

Presuppose that $G$ is acquired from two different subsequences $\{A_{x_1}, \ldots, A_{x_m}\}$ and $\{A_{y_1}, \ldots, A_{y_h}\}$, namely

$$G = \prod_{i=1}^{m} A_{x_i} = A_{x_1} \ldots A_{x_m} = \prod_{j=1}^{h} A_{y_j} = A_{y_1} \ldots A_{y_h}.$$

Since the two subsequences are unequal, there must exist a certain element $A_z$ which does not belong to the two subsequences at one time.

Without loss of generality, let $A_z \in \{A_{x_1}, \ldots, A_{x_m}\}$ and $A_z \notin \{A_{y_1}, \ldots, A_{y_h}\}$.

By the fundamental theorem of arithmetic [14], there must exist a prime $q$ which is a divisor of $A_z$.

Firstly, assume that $\forall\, A_i, A_j \in \{A_1, \ldots, A_n\}$, gcd($A_i$, $A_j$) = 1, namely $A_1, \ldots, A_n$ are pairwise coprime.

Then, there does not exist a common prime divisor between any two elements, which manifests that the prime divisors of every element do not belong to any other elements.

Thus, $q$ must be a divisor of $\prod_{i=1}^{m} A_{x_i}$ but not a divisor of $\prod_{j=1}^{h} A_{y_j}$, which means that the integer $G$ has two distinct prime factorizations, and is in direct contradiction to the fundamental theorem of arithmetic.

Secondly, assume that $\exists\, A_s, A_t \in \{A_1, \ldots, A_n\}$ with gcd($A_s$, $A_t$) $\neq 1$.

According to Definition 1, $\forall\, A_r \in \{A_1, \ldots, A_n\}$ with $r \neq s$, $t$, there are

$$(A_s / \gcd(A_s, A_t)) \nmid A_r \text{ and } (A_t / \gcd(A_s, A_t)) \nmid A_r,$$

which means that both at least one divisor of $A_s$ and at least one divisor of $A_t$ are not contained in any other elements.

Let $z = s$ or $t$, and $q$ be a prime divisor of $A_z$ with $q \nmid A_r$ $\forall\, r \neq z \in [1, n]$.

Notice that from the above assignment, we know $A_z \in \{A_{x_1}, \ldots, A_{x_m}\}$ and $A_z \notin \{A_{y_1}, \ldots, A_{y_h}\}$.

Then, there are $q \mid \prod_{i=1}^{m} A_{x_i} = G$ and $q \nmid \prod_{j=1}^{h} A_{y_j} = G$, which is in direct contradiction.

In sum, the mapping between $G$ and $\{A_{x_1}, \ldots, A_{x_m}\}$ is one-to-one. □

***Definition 2:*** Let $b_1 \ldots b_n \neq 0$ be a bit string. Then $\underline{b}_i$ with $i \in [1, n]$ is called a bit shadow if it is produced by such a rule: $\underline{b}_i$ equals 0 if $b_i = 0$, 1 plus the number of successive 0-bits before $b_i$ if $b_i = 1$, or 1 plus the number of successive 0-bits before and after $b_i$ if $b_i$ is the rightmost 1-bit.

***Fact 1:*** Let $b_1 \ldots b_n \neq 0$ be a bit string. Then there is $\sum_{i=1}^{n} \underset{\cdot}{b}_n = n$.

*Proof:*

According to Definition 2, every bit of $b_1 \ldots b_n$ is considered into $\sum_{i=1}^{k} \underset{\cdot}{b}_{x_i}$, where $\underset{\cdot}{b}_{x_1}, \ldots, \underset{\cdot}{b}_{x_k}$ are 1-bit shadows in string $\underset{\cdot}{b}_1 \ldots \underset{\cdot}{b}_n$, and there is $\sum_{i=1}^{k} \underset{\cdot}{b}_{x_i} = n$.

On the other hand, there is $\sum_{j=1}^{n-k} \underset{\cdot}{b}_{y_j} = 0$, where $\underset{\cdot}{b}_{y_1}, \ldots, \underset{\cdot}{b}_{y_{n-k}}$ are 0-bit shadows.

In total, there is $\sum_{i=1}^{n} \underset{\cdot}{b}_n = n$. □

***Property 2:*** Let $\{A_1, \ldots, A_n\}$ be a coprime sequence, and $b_1 \ldots b_n \neq 0$ be a bit string. Then the mapping from $b_1 \ldots b_n$ to $G = \prod_{i=1}^{n} A_i^{\,b_i}$ is one-to-one.

*Proof:*

Let $b_1 \ldots b_n$ and $b'_1 \ldots b'_n$ be two different bit strings, and separately correspond to $\underset{\cdot}{b}_1 \ldots \underset{\cdot}{b}_n$ and $\underset{\cdot}{b}'_1 \ldots \underset{\cdot}{b}'_n$. If $\underset{\cdot}{b}_1 \ldots \underset{\cdot}{b}_n = \underset{\cdot}{b}'_1 \ldots \underset{\cdot}{b}'_n$, it is not difficult to understand $b_1 \ldots b_n = b'_1 \ldots b'_n$. So, the mapping from $b_1 \ldots b_n$ to $\underset{\cdot}{b}_1 \ldots \underset{\cdot}{b}_n$ is one-to-one.

Additionally, since $A_i^{\underset{\cdot}{b}_i}$ is not equal to 1 and contains the same prime factors as those of $A_i$, it is known from the proof of Property 1 that the mapping from $\underset{\cdot}{b}_1 \ldots \underset{\cdot}{b}_n$ to $\prod_{i=1}^{n} A_i^{\underset{\cdot}{b}_i}$ is one-to-one.

Therefore, the mapping from $b_1 \ldots b_n$ to $\prod_{i=1}^{n} A_i^{\,b_i}$ is one-to-one. □

***Definition 3**:* The secret $\ell(i)$ in the key transform of a public key cryptosystem is called a lever function, if it has the following features:

① $\ell(i)$ is an injection from the domain $\{1, \ldots, n\}$ to the codomain $\Omega \subset \{1, \ldots, \bar{M}\}$;

② the mapping between $i$ and $\ell(i)$ is established randomly without an analytical expression;

③ an attacker has to be faced with all the permutations of elements in $\Omega$ when extracting a related private key from a public key;

④ the owner of a private key only need to considers the accumulative sum of elements in $\Omega$ when recovering a related plaintext from a ciphertext.

Feature ③ and ④ make it clear that if $n$ is large enough, it is infeasible for an attacker to search all the permutations of elements in $\Omega$ exhaustively while decryption is feasible in time being polynomial in $n$. Thus, the amount of calculation on $\ell(.)$ at 'a public terminal' is large, and the amount of calculation on $\ell(.)$ at 'a private terminal' is small.

Concretely to REESSE1+, $\ell(i)$ in the transform $C_i \equiv (A_i W^{\ell(i)})^{\delta}$ (% $M$) for $i = 1, \ldots, n$ is on the position of an exponent.

***Property 3* (Indeterminacy of $\ell$(.))*:*** Let $\delta = 1$ and $C_i \equiv A_i W^{\ell(i)}$ (% $M$) for $i = 1, \ldots, n$, where $\ell(i) \in \Omega = \{5, \ldots, n + 4\}$ and $A_i \in \Lambda = \{2, \ldots, \text{Ƥ}\}$. Then $\forall\, W \in [1, \bar{M}]$ with $\|W\| \neq \bar{M}$, and $\forall\, x, y, z \in [1, n]$ with $z \neq x, y$,

① when $\ell(x) + \ell(y) = \ell(z)$, there is $\ell(x) + \|W\| + \ell(y) + \|W\| \neq \ell(z) + \|W\|$ (% $\bar{M}$);

② when $\ell(x) + \ell(y) \neq \ell(z)$, there always exist

$C_x \equiv A'_x W'^{\,\ell'(x)}$, $C_y \equiv A'_y W'^{\,\ell'(y)}$, and $C_z \equiv A'_z W'^{\,\ell'(z)}$ (% $M$)

such that $\ell'(x) + \ell'(y) \equiv \ell'(z)$ (% $\bar{M}$) with $A'_z \leq$ Ƥ.

*Proof:*

① It is easy to understand that

$$W^{\ell(x)} \equiv W^{\ell(x)+\|W\|},\ W^{\ell(y)} \equiv W^{\ell(y)+\|W\|},\ W^{\ell(z)} \equiv W^{\ell(z)+\|W\|}\ (\%\ M).$$

Due to $\|W\| \neq \bar{M}$, $2\|W\| \neq \|W\|$, and $\ell(x) + \ell(y) = \ell(z)$, it follows that $\ell(x) + \|W\| + \ell(y) + \|W\| \neq \ell(z) + \|W\|$ (% $\bar{M}$).

However, it should be noted that when $\|W\| = \bar{M}$, there is $\ell(x) + \|W\| + \ell(y) + \|W\| = \ell(z) + \|W\|$ (% $\bar{M}$).

② Let $\bar{O}_{\mathrm{d}}$ be an oracle on seeking a discrete logarithm from DLP.

Suppose that $W' \in [1, \bar{M}]$ is a generator of $(\mathbb{Z}_M^*, \cdot)$.

In light of group theories, $\forall\, A'_z \in \{2, \ldots, \text{Ƥ}\}$, the congruence

$$C_z \equiv A'_z W'^{\,\ell'(z)}\ (\%\ M)$$

has a solution. Then, $\ell'(z)$ may be taken through $\bar{O}_{\mathrm{d}}$.

$\forall\, \ell'(x) \in (0, \bar{M})$, and let $\ell'(y) \equiv \ell'(z) - \ell'(x)$ (% $\bar{M}$).

Further, from $C_x \equiv A'_x W'^{\,\ell'(x)}$ (% $M$) and $C_y \equiv A'_y W'^{\,\ell'(y)}$ (% $M$), we can obtain many distinct pairs $(A'_x, A'_y)$, where $A'_x$, $A'_y \in [1, \bar{M}]$, and $\ell'(x) + \ell'(y) \equiv \ell'(z)$ (% $\bar{M}$). □

Notice that letting $\Omega = \{5, \ldots, n + 4\}$, namely every $\ell(i) \geq 5$ makes seeking $W$ from $W^{\ell(i)} \equiv A_i^{-1} C_i$ (% $M$) face an unsolvable Galois group when $A_i$ is guessed [22], and especially when $\Omega$ is any subset containing $n$ elements of $\{1, \ldots, \bar{M}\}$, Property 3 still holds.

Assume that $\ell(x) + \ell(y) = \ell(z)$, and let $G' \equiv C_x C_y C_z^{-1}$ (% $M$). Then

$$G' \equiv C_x C_y C_z^{-1} \equiv A_x A_y A_z^{-1}\ (\%\ M),$$

namely

$$G' / M - L / A_z = (A_x A_y) / (M A_z),$$

where $L$ is an positive integer.

Due to $M > \prod_{i=1}^{n} A_i$, $A_i \geq 2$, and $n > 3$, there is

$$G' / M - L / A_z < 1 / (2^{n-3} A_z^2) < 1 / (2 A_z^2).$$

By Theorem 12.19 in Section 12.3 of [23], $L / A_z$ is a convergent of the continued fraction expansion of $G' / M$.

Property 3 illuminates that $\ell(x) + \ell(y) = \ell(z)$ is not necessary for the above discriminant. Therefore, a continued fraction attack on $C_i \equiv A_i W^{\ell(i)}$ (% $M$) according to the discriminant will be ineffectual as long as $\Lambda$ and $\Omega$ are fitly selected [24]. However, the robust $\Lambda$ and $\Omega$ will make the decryption of a ciphertext get slow. Hence, $C_i \equiv A_i W^{\ell(i)}$ (% $M$) is not an efficient key transform.

# 3 Design of the REESSE1+ Public key Cryptosystem

In essence, REESSE1+ is a prototypal cryptosystem which is used to expound some foundational concepts, ideas, and methods.

## 3.1 The Key Generation Algorithm

This algorithm is employed by a certificate authority or the owner of a key pair.

Let $p_1, \ldots, p_n$ be the first $n$ primes in the set $\mathbb{N}$, $\Lambda = \{2, 3, \ldots, 1201\}$, and $\Omega = \{5, 7, \ldots, 2n + 3\}$. Assume that đ, Đ, $T$, $S$ are four pairwise coprime integers, where đ $\in [5, 2^{16}]$, $T \geq 2^n$, and Đ contains a prime not less than $2^n$.

S1: Randomly generate a coprime sequence $\{A_1, \ldots, A_n\}$ with $A_i \in \Lambda$.

S2: Find a prime $M > (\max_{1 \leq i \leq n} A_i)^n$ making (đĐ$T$) $|\ \bar{M}$, $\gcd(S, \bar{M}) = 1$, and $\prod_{i=1}^{k} p_i^{\,e_i} \mid \bar{M}$, where $k$ meets $\prod_{i=1}^{k} e_i \geq 2^{10}$ & $p_k \approx 2n$.

S3: Pick $W, \delta \in (1, \bar{M})$ making $\gcd(\delta, \bar{M}) = 1$, $\|\delta\| =$ đĐ$T$, and $\|W\| \geq 2^{n-20}$.

S4: Compute $\alpha \leftarrow \delta^{(\delta^n + \delta W^{n-1})T}$, $\beta \leftarrow \delta^{W^n T}$ % $M$,

$\hbar \leftarrow (W\prod_{i=1}^{n} A_i)^{-\delta S}(\alpha\delta^{-1})$ % $M$.

S5: Randomly produce pairwise distinct $\ell(1), \ldots, \ell(n) \in \Omega$.

S6: Compute $C_i \leftarrow (A_i W^{\ell(i)})^{\delta}$ % $M$ for $i = 1, \ldots, n$.

At last, regard ($\{C_i\}$, $\alpha$, $\beta$) as a public key, ($\{A_i\}$, $\{\ell(i)\}$, $W$, $\delta$, $Đ$, $đ$, $\hbar$) as a private key, and ($S$, $T$, $M$) as being in common.

Notice that if REESSE1+ is pragmatized, we suggest that the set $\Omega = \{+/-5, +/-7, \ldots, +/-(2n + 3)\}$, where every sign +/– means that '+' or '–' is selected, and unknown to the public, which may bring more indeterminacy, and that $\gcd(W, Đ) > 1$, which may avoid the existence of $W^{-1}$ % $Đ$ no matter what the value of $Đ$ is.

At S3, to seek $\delta$, first let $\delta \equiv g^{\bar{M}/(đĐT)}$ (% $M$), where $g$ is a generator by Algorithm 4.80 in Section 4.6 of [25], then test $\delta$. At S4, seeking a $S$-th root to $x^S \equiv c$ (% $M$) is referred to Theorem 1 in Section 3.4.

Considering $M > (\max_{1\le i\le n} A_i)^n$ and the fact that the first $n$ primes in the set $\mathbb{N}$ can constitute a smallest coprime sequence, we can estimate $\lg M \approx 696$, 864, 1030, or 1216 when $n = 80$, 96, 112, or 128.

***Definition 4:*** Let $\{C_1, \ldots, C_n\}$ be a non-coprime sequence, and $M$ be a prime. Seeking the original $\{A_1, \ldots, A_n\}$ with $A_i \in \Lambda$, $\{\ell(1), \ldots, \ell(n)\}$ with $\ell(i) \in \Omega$, $W$, $\delta$ from $C_i \equiv (A_i W^{\ell(i)})^{\delta}$ (% $M$) for $i = 1, \ldots, n$ is called the multivariate permutation problem, shortly MPP.

## 3.2 The Encryption Algorithm

Assume that ($\{C_i\}$, $\alpha$, $\beta$) is a public key, and $b_1\ldots b_n \ne 0$ is a plaintext block or a symmetric key.

S1: Set $\bar{G} \leftarrow 1$, $k \leftarrow 0$, $i \leftarrow 1$.

S2: If $b_i = 0$, let $k \leftarrow k + 1$, $\underset{\cdot}{b}_i \leftarrow 0$;

else do $\underset{\cdot}{b}_i \leftarrow k + 1$, $k \leftarrow 0$, $\bar{G} \leftarrow \bar{G} C_i^{\underset{\cdot}{b}_i}$ % $M$.

S3: Let $i \leftarrow i + 1$.

If $i \le n$, go to S2.

S4: If $b_n = 0$, do $\underset{\cdot}{b}_{n-k} \leftarrow \underset{\cdot}{b}_{n-k} + k$, $\bar{G} \leftarrow \bar{G}(C_{n-k})^k$ % $M$.

So, the ciphertext $\bar{G} \equiv \prod_{i=1}^{n} C_i^{\underset{\cdot}{b}_i}$ (% $M$) is obtained.

Notice that $\alpha$ and $\beta$ are unuseful for the encryption.

***Definition 5:*** Let $\{C_1, \ldots, C_n\}$ be a non-coprime sequence, and $M$ be a prime. Seeking the original $b_1\ldots b_n$ from $\bar{G}_1 \equiv \prod_{i=1}^{n} C_i^{b_i}$ (% $M$) is called the (modular) subset product problem, shortly SPP.

***Definition 6:*** Let $\{C_1, \ldots, C_n\}$ be a non-coprime sequence, and $M$ be a prime. Seeking the original $\underset{\cdot}{b}_1\ldots\underset{\cdot}{b}_n$, namely $b_1\ldots b_n$ from $\bar{G} \equiv \prod_{i=1}^{n} C_i^{\underset{\cdot}{b}_i}$ (% $M$) is called the anomalous subset product problem, shortly ASPP.

## 3.3 The Decryption Algorithm

Assume that ($\{A_i\}$, $\{\ell(i)\}$, $W$, $\delta$, $Đ$, $đ$, $\hbar$) is a related private key, and $\bar{G}$ is a ciphertext.

Notice that because $\sum_{i=1}^{n} \underset{\cdot}{b}_i = n$ is even, $\sum_{i=1}^{n} \underset{\cdot}{b}_i \ell(i)$ must be even.

S1: Compute $\bar{G} \leftarrow \bar{G}^{\delta^{-1}}$ % $M$.

S2: Compute $\bar{G} \leftarrow \bar{G} W^{-2}$ % $M$.

S3: Set $b_1\ldots b_n \leftarrow 0$, $G \leftarrow \bar{G}$, $i \leftarrow 1$, $k \leftarrow 0$.

S4: If $A_i^{k+1} \mid G$, do $G \leftarrow G / A_i^{k+1}$, $b_i \leftarrow 1$, $k \leftarrow 0$;

else let $k \leftarrow k + 1$.

S5: Let $i \leftarrow i + 1$.

If $i \le n$ and $G \ne 1$, go to S4.

S6: If $k \ne 0$ and $(A_{n-k})^k \mid G$, do $G \leftarrow G / (A_{n-k})^k$.

S7: If $G \ne 1$, go to S2; else end.

So, the original plaintext block or symmetric key $b_1\ldots b_n$ is recovered.

Only if $\bar{G}$ is a true ciphertext, can this algorithm terminate normally. In decryption, $\{\ell(i)\}$, $Đ$, $đ$, and $\hbar$ are unhelpful.

## 3.4 The Digital Signature Algorithm

Assume that ($\{A_i\}$, $\{\ell(i)\}$, $W$, $\delta$, $Đ$, $đ$, $\hbar$) is a private key, $F$ is a file or message to be signed, and *hash* is a one-way compression function.

S1: Let $H \leftarrow hash(F)$, whose binary form is $b_1\ldots b_n$.

S2: Set $\underline{k} \leftarrow \delta\sum_{i=1}^{n} b_i \ell(i)$ % $\bar{M}$, $G_0 \leftarrow (\prod_{i=1}^{n} A_i^{\neg b_i})^{\delta}$ % $M$.

S3: $\forall \bar{a} \in (1, \bar{M})$ making $(đT) \nmid \bar{a}$ and $đ \nmid (WQ)$ % $\bar{M}$,

where $Q = (\bar{a}Đ + WH)\delta^{-1}$ % $\bar{M}$.

S4: Compute $R \leftarrow (Q(\delta\hbar)^{-1})^{S^{-1}} G_0^{-1}$, $\bar{U} \leftarrow (RW^{\underline{k}-\delta})^{Q}$ % $M$,

$\bar{g} \leftarrow \delta^{\bar{a}Đ}$ % $M$, $\xi \leftarrow \sum_{i=0}^{n-1} (\delta Q)^{n-1-i}(HW)^i$ % $\bar{M}$.

S5: $\forall r \in [1, đ2^{16}]$ making $đ \nmid (rUS + \xi)$ % $\bar{M}$,

where $U = \bar{U}\bar{g}^r$ % $M$.

S6: If $đ \nmid ((WQ)^{n-1} + \xi + rUS)$ % $\bar{M}$, go to S5; else end.

So, a signature ($Q$, $U$) on the file $F$ is obtained, and sent to a receiver together with $F$.

It is known from S3, S4 that $Q, R$ meet $\bar{a}Đ \equiv \delta Q - WH$ (% $\bar{M}$) and $Q \equiv (RG_0)^S \delta\hbar$ (% $M$).

It should be noted that owing to $đ \nmid \bar{a}$, $\gcd(Đ, đ) = 1$, and $đ \mid \bar{M}$, there must exist $đ \nmid (\delta Q - WH)$.

According to the double congruence theorem (see Section 3.6), in the signature algorithm we do not need

$$V = (R^{-1}W^{\delta}G_1)^{QU}\delta^{\lambda} \text{ \% } M,$$

where $G_1 = (\prod_{i=1}^{n} A_i^{b_i})^{\delta}$ % $M$, and $\lambda$ satisfies

$$\lambda S \equiv ((WQ)^{n-1} + \xi + rUS)(\delta Q - HW) \text{ (\% } \bar{M}),$$

which indicates $(đĐ) \mid \lambda$.

At S5, the probability of finding a fit $U$ is roughly $1 / đ$. Because $đ$ is a small number, $U$ can be found out at a good pace. The small $đ$, however, does not influence the security of REESSE1+ (see Section 6.3).

Let $\Delta \equiv (WQ)^{n-1} + \xi + rUS$ (% $\bar{M}$).

Due to $đ \mid \bar{M}$, if $(WQ)^{n-1} + \xi + rUS$ contains the factor $đ$, it must be contained in $\Delta$ % $\bar{M}$.

Besides, due to $đ \nmid S$ and $đ \nmid (WQ)^{n-1}$ (according to $đ \nmid (WQ)$), if we want to make $đ \mid \Delta$, there must be $đ \nmid (rUS + \xi)$ % $\bar{M}$.

Therefore, as long as every value of $r$ makes $rUS$ different, $đ \mid \Delta$ will holds after about $đ$ attempts of $r$. The algorithm can also terminate normally because after $r$ traverses the interval $[1, đ2^{16}]$, the probability of $đ \nmid \Delta$ is $(1 - 1/đ)^{đ2^{16}}$, and almost zero.

At S4, we derive $\xi$ from $\xi(\delta Q - WH) \equiv (\delta Q)^n - (WH)^n$ (% $\bar{M}$). Computing $R$ by $Q \equiv (RG_0)^S \delta\hbar$ (% $M$) may resort to Theorem 1, where $S$ meets $\gcd(S, \bar{M}) = 1$.

***Theorem 1:*** For the congruence $x^k \equiv c$ (% $M$) with $M$ prime, if $\gcd(k, \bar{M}) = 1$, every $c$ has just one $k$-th root modulo $\bar{M}$. Especially, let $\mu$ satisfy $\mu k \equiv 1$ (% $\bar{M}$), then $c^{\mu}$ % $M$ is the $k$-th root.

Further, we have Theorem 2 and 3.

***Theorem 2:*** For the congruence $x^k \equiv c$ (% $M$) with $M$ prime, if $k \mid \bar{M}$ and $\gcd(k, \bar{M} / k) = 1$, then when $c$ is an $k$-th power residue modulo $M$, and $\mu$ satisfies $\mu k \equiv 1$ (% $\bar{M}$ / $k$), $c^{\mu}$ % $M$ is an $k$-th root.

***Theorem 3:*** For the congruence $x^k \equiv c$ (% $M$) with $M$ prime, if $k \nmid \bar{M}$, let $h = \gcd(k, \bar{M})$, $m = k / h$, and $\mu$ satisfy $\mu m \equiv 1$ (% $\bar{M}$ / $h$), then $x^k \equiv c$ (% $M$) is equivalent to

$$x^h \equiv c^{\mu} \text{ (\% } M),$$

that is, the two congruences have the same set of solutions. Furthermore, the sufficient and necessary condition for either congruence to have solutions is $c^{\bar{M}/h} \equiv 1$ (% $M$).

For the proofs of Theorem 1 and 2, refer to [26], and for the proof of Theorem 3, refer to [27]. The solution which is obtained in terms of Theorem 1 or 2, and may be written as a certain power of $c$ modulo $M$ is called the trivial solution to the congruence $x^k \equiv c$ (% $M$) [27].

### 3.5 The Identity Verification Algorithm

Assume that ($\{C_i\}$, $\alpha$, $\beta$) is a related public key, and ($Q$, $U$) is a signature on the file or message $F$.

S1: Let $H \leftarrow hash(F)$, whose binary form is $b_1 \ldots b_n$.

S2: Compute $\bar{G}_1 \leftarrow \prod_{i=1}^{n} C_i^{b_i}$ % $M$.

S3: Compute $X \leftarrow (\alpha Q^{-1})^{QUT} \alpha^{Q^n}$ % $M$,

$Y \leftarrow (\bar{G}_1{}^{Q} U^{-1})^{UST} \beta^{HQ^{n-1} + H^n}$ % $M$.

S4: If $X = Y$, the identity is valid and $F$ intact;

else the identity is invalid or $F$ modified.

By running this algorithm, a verifier can judge whether a signature is genuine or fake, prevent the signatory from denying the signature, and prevent an adversary from modifying the file.

***Definition 7:*** Let $M$ be a prime. Seeking $x \in [1, \bar{M}]$ from $y \equiv x^x$ (% $M$) is called the transcendental logarithm problem, shortly TLP.

In what follows, we argue the discriminant $X \equiv Y$ (% $M$) at S4.

It is known from Section 3.1 that

$$\alpha \equiv \delta^{(\delta^n + \delta W^{n-1})T} \equiv \delta \hbar (W^{\delta} G_0 G_1)^S \text{ and } \beta \equiv \delta^{W^n T} \text{ (\% } M).$$

Let $V \equiv (R^{-1} W^{\delta} G_1)^{QU} \delta^{\lambda}$ (% $M$).

Since $\lambda$ meets $\lambda S \equiv ((WQ)^{n-1} + \xi + rUS)(\delta Q - HW)$(% $\bar{M}$), let $\lambda = k \text{đ} \text{Đ}$, where $k$ is a integer, and then

$$\begin{aligned} Q^{QU} V^S &\equiv (R G_0)^{SQU} (\delta \hbar)^{QU} (R^{-1} W^{\delta} G_1)^{QUS} \delta^{\lambda S} \\ &\equiv (W^{\delta} G_0 G_1)^{QUS} (\delta \hbar)^{QU} \delta^{\lambda S} \\ &\equiv \alpha^{QU} \delta^{((WQ)^{n-1} + \Sigma_{i=0}^{n-1} (\delta Q)^{n-1-i} (WH)^i + rUS)(\delta Q - WH)} \\ &\equiv \alpha^{QU} \delta^{\delta W^{n-1} Q^n - W^n H Q^{n-1} + (\delta Q)^n - (WH)^n + (\delta Q - WH) rUS} \\ &\equiv \alpha^{QU} \delta^{(\delta^n + \delta W^{n-1}) Q^n} \delta^{-W^n (HQ^{n-1} + H^n)} \delta^{\bar{a} \text{Đ} rUS} \text{(\%} M). \end{aligned}$$

Transposition yields

$$V^S \equiv (\alpha Q^{-1})^{QU} \delta^{(\delta^n + \delta W^{n-1}) Q^n} \delta^{-W^n (HQ^{n-1} + H^n)} \delta^{\bar{a} \text{Đ} rUS} \text{ (\%} M).$$

Therefore, we have

$$\begin{aligned} V^{ST} &\equiv (\alpha Q^{-1})^{QUT} \delta^{(\delta^n + \delta W^{n-1}) TQ^n} \delta^{-TW^n (HQ^{n-1} + H^n)} \delta^{\bar{a} \text{Đ} rUST} \\ &\equiv (\alpha Q^{-1})^{QUT} \alpha^{Q^n} \beta^{-(HQ^{n-1} + H^n)} \delta^{\bar{a} \text{Đ} rUST} \\ &\equiv X \beta^{-(HQ^{n-1} + H^n)} \delta^{\bar{a} \text{Đ} rUST} \text{ (\% } M). \end{aligned}$$

In addition,

$$\begin{aligned} U^{UT} V^T &\equiv (RW^{\underline{k} - \delta})^{QUT} (\delta^{\bar{a} \text{Đ} r})^{UT} (R^{-1} W^{\delta} G_1)^{QUT} \delta^{\lambda T} \\ &\equiv (W^{\underline{k}} G_1)^{QUT} \delta^{\bar{a} \text{Đ} rUT} \delta^{\lambda T} \\ &\equiv \bar{G}_1{}^{QUT} \delta^{\bar{a} \text{Đ} rUT} \delta^{k \text{đ} \text{Đ} T} \\ &\equiv \bar{G}_1{}^{QUT} \delta^{\bar{a} \text{Đ} rUT} \text{ (\% } M). \end{aligned}$$

Transposition yields

$$V^T \equiv (\bar{G}_1{}^{Q} U^{-1})^{UT} \delta^{\bar{a} \text{Đ} rUT} \text{ (\% } M).$$

Hence

$$V^{ST} \equiv (\bar{G}_1{}^{Q} U^{-1})^{UST} \delta^{\bar{a} \text{Đ} rUST} \text{ (\% } M).$$

By the double congruence theorem (Theorem 4), there is

$$\begin{aligned} V^{ST} &\equiv X \beta^{-(HQ^{n-1} + H^n)} \delta^{\bar{a} \text{Đ} rUST} \\ &\equiv (\bar{G}_1{}^{Q} U^{-1})^{UST} \delta^{\bar{a} \text{Đ} rUST} \text{ (\% } M). \end{aligned}$$

Namely, $X \equiv (\bar{G}_1{}^{Q} U^{-1})^{UST} \beta^{HQ^{n-1} + H^n} \equiv Y$ (% $M$).

### 3.6 The Double Congruence Theorem

***Theorem 4*** (***The Double Congruence Theorem***)*:* Assume that $M$ is a prime, and that $s$ and $t$ satisfying $\gcd(s, t) = 1$ are two constants, then simultaneous equations

$$\begin{cases} x^s \equiv a \text{ (\% } M) \\ x^t \equiv b \text{ (\% } M) \end{cases}$$

have the unique solution if and only if $a^t \equiv b^s$ (% $M$).

*Proof:*

Necessity:

Assume that the simultaneous equations $x^s \equiv a$ (% $M$) and $x^t \equiv b$ (% $M$) have solutions.

Let $x_0$ be a solution to the two equations, then $x_0{}^s \equiv a$ (% $M$) and $x_0{}^t \equiv b$ (% $M$).

Further, $x_0{}^{st} \equiv a^t$ (% $M$) and $x_0{}^{ts} \equiv b^s$ (% $M$) can be obtained.

Therefore, $x_0{}^{st} \equiv a^t \equiv b^s$ (% $M$).

Sufficiency:

Assume that $a^t \equiv b^s$ (% $M$).

By the greatest common divisor theorem [14], there exists a pair of integers $u$ and $v$ making $us + vt = 1$. Thus, there is

$$\begin{cases} x^{us} \equiv a^u \text{ (\% } M) \\ x^{vt} \equiv b^v \text{ (\% } M). \end{cases}$$

The above two equations multiplying yields

$$x^{us+vt} \equiv x \equiv a^u \, b^v \text{ (\% } M).$$

Furthermore, we have

$$\begin{cases} (a^u b^v)^s \equiv a^{us} b^{vs} \equiv a^{us} a^{vt} \equiv a^{us+vt} \equiv a \text{ (\% } M) \\ (a^u b^v)^t \equiv a^{ut} b^{vt} \equiv b^{us} b^{vt} \equiv b^{us+vt} \equiv b \text{ (\% } M). \end{cases}$$

Accordingly, $a^u \, b^v$ is a solution to the original simultaneous equations.

Uniqueness:

Let $x_0 \equiv a^u \, b^v$ (% $M$).

Assume that another value $x_1$ meets the equations $x^s \equiv a$ (% $M$) and $x^t \equiv b$ (% $M$) at one time.

Then, it holds that

$$x_1{}^s \equiv a \text{ (\% } M) \text{ and } x_1{}^t \equiv b \text{ (\% } M).$$

By comparison, we have $x_1{}^s \equiv x_0{}^s$ and $x_1{}^t \equiv x_0{}^t$ (% $M$). Transposing gives

$$(x_0 x_1{}^{-1})^s \equiv 1 \text{ and } (x_0 x_1{}^{-1})^t \equiv 1 \text{ (\% } M).$$

If at least one between $s$ and $t$ is relatively prime to $\bar{M}$, by Theorem 1, there must be $x_0 x_1{}^{-1} \equiv 1$ (% $M$), namely $x_0 \equiv x_1$ (% $M$).

If neither $s$ nor $t$ is coprime to $\bar{M}$, may let $k = \gcd(s, \bar{M})$, $h = \gcd(t, \bar{M})$. Then we see $\gcd(s/k, \bar{M}) = 1$ and $\gcd(t/h, \bar{M}) = 1$.

Thus, there are $(x_0 x_1{}^{-1})^k \equiv 1$ and $(x_0 x_1{}^{-1})^h \equiv 1$ (% $M$). By Theorem 3 and $\gcd(s, t) = 1$, we know $\gcd(k, h) = 1$. In terms of the group theory [22], when $\gcd(k, h) = 1$, only the element '1' belongs to two different subgroup at the same time. Therefore, $x_0 x_1{}^{-1} \equiv 1$, namely $x_1 = x_0$, and $x_0$ bears uniqueness.

To sum up, we prove Theorem 4. □

### 3.7 Characteristics of REESSE1+

REESSE1+ owns the following characteristics compared with classical MH, RSA, and ElGamal cryptosystems.

- The security of REESSE1+ is not based on a single problem, but on the four problems: MPP, ASPP, TLP, and PRFP. Hence, it is a multiproblem public key cryptosystem.
- The key transform $C_i \equiv (A_i W^{\ell(i)})^\delta$ (% $M$) for $i = 1, \ldots, n$ contains $2n+2$ unknown variables, and each equation contains four almost independent variables. Hence, REESSE1+ is multivariate.
- If any of $A_i$, $W$, and $\ell(i)$ is determined, the relation between the two remainders is still nonlinear, and thus there is very complicated nonlinear relations among $A_i$, $W$, and $\ell(i)$.
- The indeterminacy of $\ell(.)$ with $\delta = 1$. If $C_i$ and $W$ are determined, $A_i$ and $\ell(i)$ cannot be determined, and even have no one-to-one relation when $W$ is a non-generator. If $C_i$ and $A_i$ are determined, $W$ and $\ell(i)$ cannot be determined, and also have no one-to-one relation for $\gcd(\ell(i), \bar{M}) > 1$.
- The insufficiency of the mapping. A private key includes $\{A_i\}$, $\{\ell(i)\}$, $W$, $\delta$ etc, but there is only a dominant mapping from $\{A_i\}$ to $\{C_i\}$, and thus the invertibility of the transform function is poor.
- Because combinations among multiple variables may bring different hardnesses, REESSE1+ is a self-improvable cryptosystem while its main architecture remains unchanged.

### 3.8 Correctness of the Decryption Algorithm

Since $(\mathbb{Z}^*_M, \cdot)$ is an Abelian, namely commutative group, $\forall \underline{k} \in [1, \bar{M}]$, there is

$$W^{\underline{k}}(W^{-1})^{\underline{k}} \equiv W^{\underline{k}} W^{-\underline{k}} \equiv 1 \ (\% \ M).$$

Let $b_1 \ldots b_n$ be an $n$-bit plaintext.

It is known from Section 3.2 that $\bar{G} \equiv \prod_{i=1}^{n} C_i^{\underline{b}_i}$ (% $M$), where $\underline{b}_i$ means what the algorithm shows, and $C_i \equiv (A_i W^{\ell(i)})^\delta$ % $M$.

Let $G \equiv \prod_{i=1}^{n} A_i^{\underline{b}_i}$ (% $M$), and $\underline{k} = \sum_{i=1}^{n} \ell(i) \underline{b}_i$.

Then, ***we need to prove*** that $\bar{G}^{\delta^{-1}}(W^{-1})^{\underline{k}} \equiv G$ (% $M$).

*Proof:*

According to the key generator and the encryption algorithm, there is

$$\begin{aligned}\bar{G} &\equiv \prod_{i=1}^{n} C_i^{\underline{b}_i} \equiv \prod_{i=1}^{n} ((A_i W^{\ell(i)})^\delta)^{\underline{b}_i} \\ &\equiv W^{(\sum_{i=1}^{n} \underline{b}_i \ell(i))\delta} \prod_{i=1}^{n} (A_i)^{\delta \underline{b}_i} \\ &\equiv W^{\underline{k}\delta} (\prod_{i=1}^{n} A_i^{\underline{b}_i})^\delta \ (\% \ M).\end{aligned}$$

Further, raising either side of the above equation to the $\delta^{-1}$-th yields

$$\begin{aligned}\bar{G}^{\delta^{-1}} &\equiv (W^{\underline{k}\delta}(\prod_{i=1}^{n} A_i^{\underline{b}_i})^\delta)^{\delta^{-1}} \\ &\equiv W^{\underline{k}} \prod_{i=1}^{n} A_i^{\underline{b}_i} \ (\% \ M).\end{aligned}$$

Multiplying either side of the just above equation by $(W^{-1})^{\underline{k}}$ yields

$$\begin{aligned}\bar{G}^{\delta^{-1}}(W^{-1})^{\underline{k}} &\equiv W^{\underline{k}} \prod_{i=1}^{n} A_i^{\underline{b}_i} (W^{-1})^{\underline{k}} \\ &\equiv W^{\underline{k}} \prod_{i=1}^{n} A_i^{\underline{b}_i} (W^{\underline{k}})^{-1} \\ &\equiv \prod_{i=1}^{n} A_i^{\underline{b}_i} \equiv G \ (\% \ M).\end{aligned}$$

Clearly, the above process also gives a method of seeking $G$ meantime. □

Notice that in practice, $b_1 \ldots b_n$ is unknowable in advance, so we have no way to directly compute $\underline{k}$. However, because the range of $\underline{k} \in (5n, n(2n + 3))$ is very narrow, we may search $\underline{k}$ heuristically by multiplying $W^{-2}$, and verify whether $G = 1$ after it is divided exactly by some $A_i^{\underline{b}_i}$. It is known from Section 3.3 that the original $b_1 \ldots b_n$ is acquired at the same time the condition $G = 1$ is satisfied.

### 3.9 Uniqueness of a Plaintext Solution to a Ciphertext

Because $\{C_1, \ldots, C_n\}$ is a non-coprime sequence, the mapping from $\prod_{i=1}^{n} C_i^{\underline{b}_i}$ % $M$ to $\bar{G}$ (see Section 3.2) is theoretically many-to-one. It might possibly result in the nonuniqueness of a plaintext solution $b_1 \ldots b_n$ when $\bar{G}$ is being unveiled.

Suppose that the ciphertext $\bar{G}$ can be obtained from two different anomalous subset products corresponding to $b_1 \ldots b_n$ and $b'_1 \ldots b'_n$ respectively. Then,

$$\bar{G} \equiv \prod_{i=1}^{n} C_i^{\underline{b}_i} \equiv \prod_{i=1}^{n} C_i^{\underline{b}'_i} \ (\% \ M).$$

That is,

$$\prod_{i=1}^{n} (A_i W^{\ell(i)})^{\delta \underline{b}_i} \equiv \prod_{i=1}^{n} (A_i W^{\ell(i)})^{\delta \underline{b}'_i} \ (\% \ M).$$

Further, there is

$$W^{\underline{k}\delta} \prod_{i=1}^{n} (A_i)^{\delta \underline{b}_i} \equiv W^{\underline{k}'\delta} \prod_{i=1}^{n} (A_i)^{\delta \underline{b}'_i} \ (\% \ M),$$

where $\underline{k} = \sum_{i=1}^{n} \underline{b}_i \ell(i)$, and $\underline{k}' = \sum_{i=1}^{n} \underline{b}'_i \ell(i)$ % $\bar{M}$.

Raising either side of the above congruence to the $\delta^{-1}$-th power yields

$$W^{\underline{k}} \prod_{i=1}^{n} A_i^{\underline{b}_i} \equiv W^{\underline{k}'} \prod_{i=1}^{n} A_i^{\underline{b}'_i} \ (\% \ M).$$

Without loss of generality, let $\underline{k} \geq \underline{k}'$. Because $(\mathbb{Z}^*_M, \cdot)$ is an Abelian group, there is

$$W^{\underline{k}-\underline{k}'} \equiv \prod_{i=1}^{n} A_i^{\underline{b}'_i} (\prod_{i=1}^{n} A_i^{\underline{b}_i})^{-1} \ (\% \ M).$$

Let $\theta \equiv \prod_{i=1}^{n} A_i^{\underline{b}'_i} (\prod_{i=1}^{n} A_i^{\underline{b}_i})^{-1}$ (% $M$), namely $\theta \equiv W^{\underline{k}-\underline{k}'}$ (% $M$).

The above congruence signifies that when the plaintext $b_1 \ldots b_n$ is not unique, the value of $W$ must be relevant to $\theta$. The contrapositive assertion equivalent to it is that if the value of $W$ is irrelevant to $\theta$, $b_1 \ldots b_n$ will be unique. Thus, we need to consider the probability that $W$ takes a value relevant to $\theta$.

If an adversary tries to attack an 80-bit symmetric key through exhaustive search, and a computer can verify trillion values per second, it will take 38334 years for the adversary to verify all the potential values. Hence, currently 80 bits are quite enough for the security of a symmetric key.

$b_1 \ldots b_n$ contains $n$ bits which indicates $\prod_{i=1}^{n} A_i^{\underline{b}_i}$ has $2^n$ potential values, and thus the number of potential values of $\theta$ is at most $2^n \times 2^n$. Notice that because $A_1^{-1}, \ldots, A_n^{-1}$ are not necessarily coprime, some values of $\theta$ may possibly occur repeatedly.

Because $|\underline{k} - \underline{k}'| \leq n(2n + 3) - 5n \leq 32512 \approx 2^{15}$ with $n \leq 128$, and $W$ has at most $2^{15}$ solutions to every $\theta$, the probability that $W$ takes a value relevant to $\theta$ is at most $2^{15} 2^{2n} / M$. When $n \geq 80$, there is

$$2^{15} 2^{2n} / M \leq 2^{175} / 2^{696} = 1 / 2^{521}$$

which is close to zero. The probability will further decrease when $W$ is a prime since the solutions to $\theta$ lean toward being composite integers averagely.

In addition, if please, resorting to $\sum_{i=1}^{n} b_i = n$, you may exclude some unoriginal plaintext solutions.

# 4 Security Analysis of the Key Transform

We analyze the exact security of the REESSE1+ key transform $C_i \equiv (A_i W^{\ell(i)})^\delta$ (% $M$) for $i = 1 \ldots, n$, where $W, \delta \in [1, \bar{M}]$, $\ell(i) \in \Omega = \{5, 7, \ldots, 2n + 3\}$, and $A_i \in \Lambda = \{2, 3, \ldots, 1201\}$.

We know that when $n$ = 80, 96, 112, or 128, there is $\lceil \lg M \rceil \approx$ 696, 864, 1030, or 1216. In this case, IFP and DLP can almost be solved in tolerable time, and LSSP with $D \approx n / \lceil \lg M \rceil$ can also be solved in tolerable time [11][28]. In addition, because the root finding problem (RFP) $y \equiv x^k$ (% $M$) may be converted into a linear congruence through a discrete logarithm, RFP can also be solved in tolerable time when DLP can be solved in tolerable time.

'Tolerable time' indicates that the running time of an algorithm for solving a problem may be accepted by a user when the time dominant parameter is relatively small. For example, when $n$ = 80, the time $O(2^{n/2})$ is tolerable, and when $\lceil \lg M \rceil = 384$, $O(L_M[1/3, 1.923]) = 2^{56}$ is also tolerable [29].

A public key may be regarded as the cipher of a private key. Since a ciphertext is the mutual effect of a public key and a plaintext, averagely the ciphertext has no direct help to inferring the private key.

***Definition 8**:* Let $A$ and $B$ be two computational problems. $A$ is said to reduce to $B$ in polynomial time, written as $A \le_T^P B$, if there is an algorithm for solving $A$ which calls, as a subroutine, a hypothetical algorithm for solving $B$, and runs in polynomial time, excluding the time of the algorithm for $B$ [25][28].

The hypothetical algorithm for solving $B$ is called an oracle. It is easy to understand that no matter what the running time of the oracle is, it does not influence the result of the comparison.

$A \le_T^P B$ means that the difficulty of $A$ is not greater than that of $B$, namely the running time of the fastest algorithm for $A$ is not greater than that of the fastest algorithm for $B$ when all polynomial times are treated as being pairwise equivalent. Concretely speaking, if $A$ cannot be solved in polynomial or subexponential time, $B$ cannot also be solved in corresponding polynomial or subexponential time; and if $B$ can be solved in polynomial or subexponential time, $A$ can also be solved in corresponding polynomial or subexponential time.

***Definition 9**:* Let $A$ and $B$ be two computational problems. If $A \le_T^P B$ and $B \le_T^P A$, then $A$ and $B$ are said to be computationally equivalent, written as $A =_T^P B$ [25][28].

$A =_T^P B$ means that either if $A$ is a hardness of a certain complexity on condition that the dominant variable approaches a large number, $B$ is also a hardness of the same complexity on the identical condition; or $A$, $B$ both can be solved in linear or polynomial time.

Definition 8 and 9 suggest a reductive proof method called polynomial time Turing reduction (PTR) [25]. Provable security by PTR is substantially relative and asymptotic just as a one-way function is. Relative security implies that the security of a cryptosystem based on a problem is comparative, but not absolute. Asymptotic security implies that even if a cryptosystem based on a problem is proven to be secure, it is practically secure only on condition that the dominant parameter is large enough.

Naturally, we will enquire whether $A <_T^P B$ exists or not. The definition of $A <_T^P B$ may possibly be given theoretically, but the proof of $A <_T^P B$ is not easy in practice.

Let $\hat{H}(y = f(x))$ represent the complexity or hardness of solving the problem $y = f(x)$ for $x$ [30].

## 4.1 MPP Is at Least Equivalent to DLP

Definition 4 refers to $C_i \equiv (A_i W^{\ell(i)})^\delta$ (% $M$) for $i = 1, \ldots, n$ as MPP. It has the following property.

***Property 4:*** MPP is computationally at least equivalent to DLP in the same prime field.

*Proof:*

Firstly, systematically consider $C_i \equiv (A_i W^{\ell(i)})^\delta$ (% $M$) for $i = 1, \ldots, n$.

Assume that $g_i \equiv A_i W^{\ell(i)}$ (% $M$) for each $i$ is a constant.

Let

$$g_i \equiv g^{x_i} \text{ (\% } M), \text{ and } z_i \equiv \delta x_i \text{ (\% } \bar{M}),$$

where $g \in \mathbb{Z}_M^*$ be a generator.

Then, there is

$$C_i \equiv g_i^\delta \equiv g^{\delta x_i} \text{ (\% } M) \text{ for } i = 1, \ldots, n.$$

Again let $\delta x_i \equiv z_i$ (% $\bar{M}$). Then

$$C_i \equiv g^{z_i} \text{ (\% } M) \text{ for } i = 1, \ldots, n.$$

The above expression corresponds to the fact that in the ElGamal cryptosystem with many users sharing a modulus and a generator, user 1 acquires the private key $z_1$ and the public key $C_1$, …, user $n$ acquires the private key $z_n$ and the public key $C_n$. It is well known that in this case, attack of adversaries is still faced with DLP, namely seeking $z_i$ from $C_i \equiv g^{z_i}$ (% $M$) for $i = 1, \ldots, n$ is equivalent to DLP [25].

Thus, when every $g_i$ is weakened to a constant, seeking $\delta$ from $C_i \equiv g_i^\delta$ (% $M$) for $i = 1, \ldots, n$ is equivalent to DLP, which indicates that when every $g_i$ is not a constant, seeking $g_i$ and $\delta$ from $C_i \equiv g_i^\delta$ (% $M$) for $i = 1, \ldots, n$ is at least equivalent to DLP.

Secondly, singly consider a certain $C_i$, where the subscript $i$ is designated.

Assume that $\bar{O}_m(C_i, M, \underline{R})$ is an oracle on solving $C_i \equiv g_i^\delta$ (% $M$) for $g_i$ and $\delta$, where $i$ is in $\{1, \ldots, n\}$, and $\underline{R}$ is a constraint on $g_i$ such that the original $g_i$ can be found.

Let $y \equiv g^x$ (% $M$) be of DLP. Then, by calling $\bar{O}_m(y, M, g)$, $x$ can be obtained.

According to Definition 8, there is

$$\hat{H}(y \equiv g^x \text{ (\% } M)) \le_T^P \hat{H}(C_i \equiv g_i^\delta \text{ (\% } M)),$$

which means that when only a certain $g_i$ is known, seeking $g_i$ and $\delta$ from $C_i \equiv g_i^\delta$ (% $M$) is at least equivalent to DLP.

Integrally, seeking the original $\{A_i\}$, $\{\ell(i)\}$, $W$, and $\delta$ from $C_i \equiv (A_i W^{\ell(i)})^\delta$ (% $M$) for $i = 1, \ldots, n$ is computationally at least equivalent to DLP in the same prime field. □

Further, the following analysis will incline people to believe that MPP is harder than DLP.

## 4.2 Attacks by Interaction of the Key Transform Items

Every $\ell(i) \in \{5, 7, \ldots, 2n + 3\}$ and every $A_i \in \{2, 3, \ldots, 1201\}$ are the thinness of $C_i \equiv (A_i W^{\ell(i)})^\delta$ (% $M$). Naturally adversaries will adopt combinational attack measures around the thinness.

### 4.2.1 Eliminating W through $\ell(x_1)+\ell(x_2) = \ell(y_1) + \ell(y_2)$

$\forall\, x_1, x_2, y_1, y_2 \in [1, n]$, assume that there is $\ell(x_1) + \ell(x_2) = \ell(y_1) + \ell(y_2)$.

Let $G_z \equiv C_{x_1}C_{x_2}(C_{y_1}C_{y_2})^{-1}$ (% $M$), namely

$$G_z \equiv (A_{x_1}A_{x_2}(A_{y_1}A_{y_2})^{-1})^{\delta} \text{ (\% } M).$$

If the adversaries divine the values of $A_{x_1}$, $A_{x_2}$, $A_{y_1}$, and $A_{y_2}$, and compute $u$, $v_{x_1}$, $v_{x_2}$, $v_{y_1}$, $v_{y_2}$ in time of at least $L_M[1/3, 1.923]$ such that

$$G_z \equiv g^u,\ A_{x_1} \equiv g^{v_{x_1}},\ A_{x_2} \equiv g^{v_{x_2}},\ A_{y_1} \equiv g^{v_{y_1}},\ A_{y_2} \equiv g^{v_{y_2}} \text{ (\% } M),$$

where $g$ is a generator of $(\mathbb{Z}^*_M, \cdot)$, then

$$u \equiv (v_{x_1} + v_{x_2} - v_{y_1} - v_{y_2})\delta \text{ (\% } \bar{M}).$$

If $\gcd(v_{x_1} + v_{x_2} - v_{y_1} - v_{y_2}, \bar{M}) \mid u$, the congruence in $\delta$ has solutions. Because each of $A_{x_1}$, $A_{x_2}$, $A_{y_1}$, $A_{y_2}$ may traverse the interval $\Lambda$, the subscripts $x_1$, $x_2$, $y_1$, $y_2$ are unfixed, and the congruence may have $n$ solutions, the number of potential values of $\delta$ is about $n^5\|\Lambda\|^4$.

In succession, the most effectual approach seeking $W$ is that for every $i$, divine $A_i$ and $\ell(i)$, find $V_i$ by $C_i \equiv (A_i W^{\ell(i)})^{\delta}$ (% $M$), namely the value set of $W$, and if there exists $W_1 \in V_1$, …, $W_n \in V_n$ being equal pairwise, the divination of $\delta$, $\{A_i\}$, and $\{\ell(i)\}$ is thought right. Notice that to avoid seeking $\ell(i)$-th roots, may let $W = g^{\mu}$ % $M$.

Due to $\prod_{i=1}^{k} p_i^{e_i} \mid \bar{M}$, where $k$ meets $p_k \approx 2n$, there is $\ell(i) \mid \bar{M}$, and the size of every $V_i$ is about $n\|\Omega\|\|\Lambda\|$.

In summary, the running time of the above attack is at least

$$\mathcal{T} = n\|\Lambda\|L_M[1/3, 1.923] + (2n^5\|\Lambda\|^4)2\lceil \lg M\rceil^2 + (2n^5\|\Lambda\|^4)(n\|\Omega\|\|\Lambda\|)n(2\lceil \lg M\rceil^2).$$

When $n = 80$ with $\lceil \lg M\rceil \approx 696$, $\mathcal{T} = 2^{110} > 2^n$.

When $n = 96$ with $\lceil \lg M\rceil \approx 864$, $\mathcal{T} = 2^{115} > 2^n$.

When $n = 112$ with $\lceil \lg M\rceil \approx 1030$, $\mathcal{T} = 2^{125} > 2^n$.

When $n = 128$ with $\lceil \lg M\rceil \approx 1216$, $\mathcal{T} = 2^{129} \approx 2^n$.

Therefore, $\mathcal{T}$ is not less than a quantity of time exponential in $n$.

Clearly, the running time of attack by eliminating $W$ through $\ell(x_1) + \ell(x_2) + \ell(x_3) = \ell(y_1)$ is the same as that of the attack by eliminating $W$ through $\ell(x_1) + \ell(x_2) = \ell(y_1) + \ell(y_2)$.

### 4.2.2 Eliminating $W$ through the $\|W\|$-th Power

Due to $\lceil \lg M\rceil \approx 696$, 864, 1030, or 1216, $\bar{M}$ can be factorized in tolerable time. Again due to

$$\prod_{i=1}^{k} p_i^{e_i} \mid \bar{M} \text{ and } \prod_{i=1}^{k} e_i \ge 2^{10},$$

where $k$ meets $p_k \approx 2n$, $\|W\|$ can be divined in the running time of about $2^{10}$.

Raising either side of $C_i \equiv (A_i W^{\ell(i)})^{\delta}$ % $M$ to the $\|W\|$-th power yields

$$C_i^{\|W\|} \equiv (A_i)^{\delta\|W\|} \text{ \% } M.$$

Let $C_i \equiv g^{u_i}$ (% $M$), and $A_i \equiv g^{v_i}$ (% $M$), where $g$ is a generator of $(\mathbb{Z}^*_M, \cdot)$. Then

$$u_i\|W\| \equiv v_i\|W\|\delta \text{ (\% } \bar{M})$$

for $i = 1, …, n$. Notice that $u_i \neq v_i\delta$ (% $\bar{M}$).

The above congruence looks to be the MH transform [8]. Actually, $\{v_1\|W\|, …, v_n\|W\|\}$ is not a super increasing sequence, and moreover there is not necessarily $\lg(u_i\|W\|) = \lg \bar{M}$.

Because $v_i\|W\| \in [1, \bar{M}]$ is stochastic, the inverse $\delta^{-1}$ % $\bar{M}$ not need be close to the minimum

$$\bar{M}/(u_i\|W\|),\ 2\bar{M}/(u_i\|W\|),\ …, \text{ or } (u_i\|W\| - 1)\bar{M}/(u_i\|W\|).$$

Namely $\delta^{-1}$ may lie at any integral position of the interval

$$[k\bar{M}/(u_i\|W\|),\ (k + 1)\bar{M}/(u_i\|W\|)],$$

where $k = 0, 1, …, u_i\|W\| - 1$, which illustrates the accumulation points of minima do not exist. Further observing, in this case, when $i$ traverses the interval $[2, n]$, the number of intersections of the intervals including $\delta^{-1}$ is likely $\max_{2 \le i \le n} \{u_i\|W\|\}$ which is promisingly close to $\bar{M}$. Therefore, the Shamir attack by the accumulation point of minima is fully ineffectual [9].

Even if find out $\delta^{-1}$ by the Shamir attack method, because each of $v_i$ has $\|W\|$ solutions, the number of potential sequences $\{g^{v_1}, …, g^{v_n}\}$ is up to $\|W\|^n$. Because of needing to verify whether $\{g^{v_1}, …, g^{v_n}\}$ is a coprime sequence for each different sequence $\{v_1, …, v_n\}$, the number of coprime sequences is in proportion to $\|W\|^n$. Hence, the initial $\{A_1, …, A_n\}$ cannot be determined in polynomial time. Further, the value of $W$ cannot be computed, and the values of $\|W\|$ and $\delta^{-1}$ cannot be verified in polynomial time, which indicates that MPP can also be resistant to the attack by the accumulation point of minima.

Additionally, the adversaries may divine value of $A_i$ in running time of about $\|\Lambda\|$, where $i \in [1, n]$, and compute $\delta$ by $u_i\|W\| \equiv v_i\|W\|\delta$ (% $\bar{M}$). However, because of $\|W\| \mid \bar{M}$, the equation will have $\|W\|$ solutions. Therefore, the running time of finding the original $\delta$ is at least

$$\begin{aligned}\mathcal{T} &= n\|\Lambda\|L_M[1/3, 1.923] + 2^{10}\|\Lambda\|\|W\| \\ &= n\|\Lambda\|L_M[1/3, 1.923] + 2^{10}\|\Lambda\|2^{n-20} \\ &\approx n\|\Lambda\|L_M[1/3, 1.923] + 2^n > 2^n.\end{aligned}$$

It is at least exponential in $n$ when $80 \le n \le 128$.

Again, it is infeasible to separate $\hbar$, $\delta$, $W$, and $\prod_{i=1}^{n} A_i$ distinctly from $\alpha \equiv \hbar\,\delta(W\prod_{i=1}^{n} A_i)^{-\delta S}$ (% $M$), and the time complexity of seeking $\delta$, $W$ from $\alpha \equiv \delta^{(\delta^n + \delta W^{n-1})T}$ and $\beta \equiv \delta^{W^n T}$ (% $M$) will be at least $O(2^n)$ (see Section 6.3.3), so even if the three equations are considered simultaneously, it is also impossible to determine the values of the four variables almost independent.

In summary, the time complexity of inferring a related private key from a public key is at least $O(2^n)$.

## 4.3 Attack by a Certain Single $C_i$

Assume that there is only a solitary $C_i = (A_i W^{\ell(i)})^{\delta}$ % $M$ — $i = 1$ for example, and other $C_i$'s ($i = 2, …, n$) are unknown for attackers.

Through divining $A_1 \in \Lambda$ and $\ell(1) \in \Omega$, the parameters $W$ and $\delta \in (1, \bar{M})$ can be computed. Thus, the number of solution $(A_1, \ell(1), W, \delta)$ will be up to $\|\Omega\|\|\Lambda\|\bar{M}^2 > 2^n$, which manifests that the original $(A_1, \ell(1), W, \delta)$ cannot be determined in time being subexponential in $n$ [30].

Evidently, if $g_1 \equiv A_1 W^{\ell(1)}$ (% $M$) is a constant, solving $C_1 = g_1^{\delta}$ % $M$ for $\delta$ is equivalent to DLP. Factually, $g_1$ is not a constant. At present, seeking the original $g_1$, $\delta$ will take at least $O(M) > O(2^n)$ steps.

# 5 Security Analysis of the Encryption Algorithm

The security of the encryption algorithm is namely the security of a REESSE1+ ciphertect.

## 5.1 ASPP Is at Least Equivalent to DLP

Definition 5 refers to $\bar{G}_1 \equiv \prod_{i=1}^{n} C_i^{b_i}$ (% $M$) as SPP. It has the following property.

***Property 5:*** SPP is computationally at least equivalent to DLP in the same prime field.

*Proof:*

Let $\bar{G}_1 \equiv \prod_{i=1}^{n} C_i^{b_i}$ (% $M$), where $b_1 \ldots b_n$ is a plaintext block.

Especially, define $\bar{G}_1 \equiv \prod_{i=1}^{n} C^{2^{n-i} b_i} \equiv \prod_{i=1}^{n} (C^{2^{n-i}})^{b_i}$ (% $M$) when $C_1 = \ldots = C_n = C$.

Obviously, $\prod_{i=1}^{n} C_i^{b_i} = L\,M + \bar{G}_1$. Owing to $L \in [1, \bar{M}]$, deriving the non-modular product $\prod_{i=1}^{n} C_i^{b_i}$ from $\bar{G}_1$ is infeasible, which means inferring $b_1 \ldots b_n$ from $\bar{G}_1$ is not a factorization problem.

Assume that $\bar{O}_s(\bar{G}_1, C_1, \ldots, C_n, M)$ is an oracle on solving $\bar{G}_1 \equiv \prod_{i=1}^{n} C_i^{b_i}$ (% $M$) for $b_1 \ldots b_n$.

Let $y \equiv g^x$ (% $M$) be of DLP, where $g$ is a generator of $(\mathbb{Z}_M^*, \cdot)$, and the binary form of $x$ is $b'_1 \ldots b'_n$, namely $y \equiv \prod_{i=1}^{n} (g^{2^{n-i}})^{b'_i}$ (% $M$).

Then, by calling $\bar{O}_s(y, g^{2^{n-1}}, \ldots, g, M)$, $x$ namely $b'_1 \ldots b'_n$ can be found.

By Definition 8, there is

$$\hat{H}(y \equiv g^x \text{ (\% } M)) \le_T^P \hat{H}(\bar{G}_1 \equiv \textstyle\prod_{i=1}^{n} C_i^{b_i} \text{ (\% } M)),$$

namely SPP is at least equivalent to DLP in the same prime field in complexity. □

Definition 6 refers to $\bar{G} \equiv \prod_{i=1}^{n} C_i^{ƀ_i}$ (% $M$) as ASPP. It has a similar property.

***Property 6:*** ASPP is computationally at least equivalent to DLP in the same prime field.

*Proof:*

Assume that $\bar{O}_a(\bar{G}, C_1, \ldots, C_n, M)$ is an oracle on solving $\bar{G} \equiv \prod_{i=1}^{n} C_i^{ƀ_i}$ (% $M$) for $ƀ_1 \ldots ƀ_n$, where $ƀ_1 \ldots ƀ_n$ is the bit shadow string of $b_1 \ldots b_n$.

Especially, define $\bar{G} \equiv \prod_{i=1}^{n} C^{n^{n-i} ƀ_i} \equiv \prod_{i=1}^{n} (C^{n^{n-i}})^{ƀ_i}$ (% $M$) with the stipulation $ƀ_i < n$ (namely that $b_1 \ldots b_n$ contains at least two nonzero bits) when $C_1 = \ldots = C_n = C$.

Let $\bar{G}_1 \equiv \prod_{i=1}^{n} C_i^{b_i}$ (% $M$) be of SPP.

Due to $0 \le b_i \le ƀ_i$, by calling $\bar{O}_a(\bar{G}_1, C_1, \ldots, C_n, M)$, $b_1 \ldots b_n$ can be found.

By Definition 8, there is

$$\hat{H}(\bar{G}_1 \equiv \textstyle\prod_{i=1}^{n} C_i^{b_i} \text{ (\% } M)) \le_T^P \hat{H}(\bar{G} \equiv \textstyle\prod_{i=1}^{n} C_i^{ƀ_i} \text{ (\% } M)).$$

Further by transitivity, there is

$$\hat{H}(y \equiv g^x \text{ (\% } M)) \le_T^P \hat{H}(\bar{G} \equiv \textstyle\prod_{i=1}^{n} C_i^{ƀ_i} \text{ (\% } M)),$$

namely ASPP is at least equivalent to DLP in the same prime field in complexity. □

It is very interesting whether ASPP is harder than DLP or not. We can find some positive evidence.

Piece 1 of evidence:

Observe an extreme case.

Assume that $C_1 = \ldots = C_n = C$, then we have

$$\bar{G} \equiv \textstyle\prod_{i=1}^{n} C^{ƀ_i n^{n-i}} \text{ (\% } M)$$

which may be written as

$$\bar{G} \equiv C^{\sum_{i=1}^{n} ƀ_i n^{n-i}} \text{ (\% } M).$$

Let

$$z = \textstyle\sum_{i=1}^{n} ƀ_i n^{n-i}.$$

Correspondingly,

$$\bar{G} \equiv C^z \text{ (\% } M),$$

which is a form of DLP.

However, when $C_1, \ldots, C_n$ are generated, we can check $C_1, \ldots, C_n$ to prevent any two elements in $\{C_1, \ldots, C_n\}$ from being equal. Therefore, in practice, ASPP cannot be reduced to DLP in any time.

Piece 2 of evidence:

Assume that DLP can be solved in tolerable subexponential time.

When DLP can be solved in tolerable time, $\bar{M}$ can also be factorized [14][25], so a generator can be found through the algorithm 4.80 in Section 4.6 of [25].

Let $C_1 \equiv g^{u_1}$ (% $M$), …, $C_n \equiv g^{u_n}$ (% $M$), $\bar{G} \equiv g^v$ (% $M$), where $g$ is a generator of $(\mathbb{Z}_M^*, \cdot)$.

Then, solving $\bar{G} \equiv \prod_{i=1}^{n} C_i^{ƀ_i}$ (% $M$) for $ƀ_1 \ldots ƀ_n$ is equivalent to solving

$$ƀ_1 u_1 + \ldots + ƀ_n u_n \equiv v \text{ (\% } \bar{M})$$

which is called the anomalous subset sum problem (ASSP) due to $ƀ_i \ge b_i$, and computationally at least equivalent to SSP.

It has been proven that SSP is NP-complete in its feasibility recognition form, and the computational version of SSP with the sufficiently large length and density is NP-hard [4][25]. Hence, solving ASSP is at least NP-hard. Additionally, the density relevant to ASSP is far greater than 1 because of $ƀ_i \ge b_i$ (see Appendix B), which indicates that the $L^3$ lattice base reduction attack on ASSP will be ineffectual (see Section 5.2). These two points illustrate that even although DLP can be solved, $ƀ_1 \ldots ƀ_n$, namely $b_1 \ldots b_n$ cannot be found yet in polynomial time.

The above two pieces of evidence incline us to believe that ASPP is harder than DLP.

By the way, LSSP will degenerate to a polynomial time problem from NPC [11][31].

## 5.2 ASPP Can Resist $L^3$ Lattice Base Reduction

It is known from Section 3.2 that the ciphertext $\bar{G} \equiv \prod_{i=1}^{n} C_i^{ƀ_i}$ (% $M$).

Still let $C_1 \equiv g^{u_1}$ (% $M$), …, $C_n \equiv g^{u_n}$ (% $M$), $\bar{G} \equiv g^v$ (% $M$), where $g$ is a generator of $(\mathbb{Z}_M^*, \cdot)$ randomly selected.

Then, seeking $ƀ_1 \ldots ƀ_n$ from $\bar{G}$ is equivalent to solving the congruence

$$u_1 ƀ_1 + \ldots + u_n ƀ_n \equiv v \text{ (\% } \bar{M}), \qquad (1)$$

where $v$ may be substituted with $v + k\bar{M}$ with $k \in [0, n-1]$ [32]. $\{u_1, \ldots, u_n\}$ is called a compact sequence due to $ƀ_i \in [0, n]$ and $n > 1$.

Recall [10] and [11]. Let $\{a_1, \ldots, a_n\}$ be a positive integer sequence, $\hat{e} = \langle \dot{e}_1, \ldots, \dot{e}_n, 0 \rangle$ with $\dot{e}_i \in [0, 1]$ be the solution vector, $s = \sum_{i=1}^{n} a_i \dot{e}_i$, and $t = \sum_{i=1}^{n} a_i$.

In [10], there are two important conditions:

$$t / n \le s \le (n-1)\, t / n, \text{ and } \|\hat{e}\|^2 \le n / 2,$$

where $\|\hat{e}\|$ denotes the distance in $l_2$-Norm of the vector $\hat{e}$,

which decides the threshold density < 0.6463.

In [11], there are similar

$$t / n \le s \le (n-1)t / n, \text{ and } \|\hat{e}\|^2 \le n / 4,$$

which decide the threshold density < 0.9408.

However, for (1), due to $0 \le b_i \le n$, the similar conditions do not hold.

It is well understood that the $L^3$ lattice base reduction algorithm is employed in cryptanalysis to find the shortest vector or approximately shortest vectors in a lattice, and hence, if a solution to SSP has a comparatively big distance, or is not unique, it will not occur in the reduced base.

Let $\mathbb{L}$ be a lattice spanned by the vectors

$$\langle 1, 0, \ldots, 0, Nu_1\rangle,$$
$$\langle 0, 1, \ldots, 0, Nu_2\rangle,$$
$$\vdots \; \vdots \qquad \vdots \; \vdots$$
$$\langle 0, 0, \ldots, 1, Nu_n\rangle,$$
$$\langle 0, 0, \ldots, 0, N(v+k\bar{M})\rangle$$

which compose a base of the lattice, where $N$ is a positive integer greater than $(n^2)^{1/2} = n$ (but not much greater, or else will influence speed of the $L^3$ reduction algorithm). Notice that because $g$ is random, $\mathbb{L}$ is also random.

Let $D$ be the determinant relevant to a matrix corresponding to the lattice base. Then, by the Guassian heurisic, the expected size of the shortest vector in $\mathbb{L}$ of $n$ + 1 dimensions lies between [15]

$$D^{1/(n+1)}((n+1)/(2\pi e))^{1/2} \text{ and } D^{1/(n+1)}((n+1)/(\pi e))^{1/2},$$

where $e \approx 2.7182818$.

In our case, there is $\lceil \lg M \rceil / (n+1) \approx 9$, and the above scope is between

$$(Nk2^{9(n+1)})^{1/(n+1)}((n+1)/(2\pi e))^{1/2}$$

and

$$(Nk2^{9(n+1)})^{1/(n+1)}((n+1)/(\pi e))^{1/2}.$$

Roughly, between

$$2^9((n+1)/2^4)^{1/2} = 2^7 n^{1/2} \text{ and } 2^7(2n)^{1/2} = 2^9((n+1)/2^3)^{1/2}.$$

For (1), the largest distance of the solution vector $\langle b_1, \ldots, b_n, 0\rangle$ is $n \in \{80, 96, 112, 128\}$, and thus it is very possible that the solution vector will not occur in the reduced base. Meanwhile, it will also be influenced by the knapsack density relevant to (1) whether the solution vector surely occurs in the reduced base.

To compute the density of the compact sequence, we extend $\{u_1, \ldots, u_n\}$ into

$$\{u_1, 2u_1, \ldots, nu_1, u_2, 2u_2, \ldots, nu_2, \ldots\ldots, u_n, 2u_n, \ldots, nu_n\}.$$

It is not difficult to understand that the length of the extend sequence is $n^2$.

The density of the compact sequence $\{u_1, \ldots, u_n\}$ is

$$D \approx n^2 / \lceil \lg M \rceil \text{ (see Appendix B).}$$

When $n = 80$ with $\lceil \lg M \rceil = 696$, $D \approx 9.19 > 2 > 1$.

When $n = 96$ with $\lceil \lg M \rceil = 864$, $D \approx 10.66 > 2 > 1$.

When $n = 112$ with $\lceil \lg M \rceil = 1030$, $D \approx 12.18 > 2 > 1$.

When $n = 128$ with $\lceil \lg M \rceil = 1216$, $D \approx 13.47 > 2 > 1$.

$D > 2$ indicates that a great many different subsets will have the identical sum, namely the solution to (1) is not unique, and the original solution is possibly not shortest for $b_i \in [0, n]$. Thus, it is very likely that the original solution does not occur in the reduced base only containing $n$ + 1 vectors.

Further, we can estimate the time cost of the $L^3$ lattice base attack.

Although SLLL, namely segment LLL in floating point arithmetic and $L^2$-FP are two of currently fast lattice base reduction algorithms [33][34], because floating point operation on integers greater than the modulus $M$ with $\lg M \ge 696$ cannot be executed directly, and even are instable under a low precision circumstance, it is inappropriate to utilize these two algorithms to find the solution vector $\langle b_1, \ldots, b_n, 0\rangle$, which manifests that the only classical $L^3$ algorithm is appropriate.

According to [25], the running time of attack on equation (1) from ASPP by the lattice base reduction algorithm is roughly

$$T \approx O(nL_M[1/3, 1.923] + n(n+1)^6(\lg M^2)^3)$$

on condition that $N$ is slightly greater than $n$.

When $n = 80$ with $\lceil \lg M \rceil = 696$, $T \approx 2^{76}$.

When $n = 96$ with $\lceil \lg M \rceil = 864$, $T \approx 2^{80}$.

When $n = 112$ with $\lceil \lg M \rceil = 1030$, $T \approx 2^{82}$.

When $n = 128$ with $\lceil \lg M \rceil = 1216$, $T \approx 2^{86}$.

However, as is pointed out in the above, owing to $D > 9 > 2 > 1$ and $b_i \in [0, n]$, it is almost impossible that the solution vector $\langle b_1, \ldots, b_n, 0\rangle$ occurs in the final reduced base, which means that attack by the $L^3$ algorithm will be unavailing.

Besides, we also see that there exists an exhaustive search attack on the plaintext block $b_1 \ldots b_n$. Clearly, the running time of such an attack is $O(2^n)$ arithmetic steps.

Hence, the security of a REESSE1+ plaintext is built on the problem $\bar{G} \equiv \prod_{i=1}^{n} C_i^{b_i} (\% M)$ which contains the trapdoor information, and means that computing an anomalous subset product from subset elements is tractable while seeking the subset elements from the product is intractable.

### 5.3 Avoid Adaptive-chosen-ciphertext Attack

Theoretically, most of public key cryptosystems may probably be faced with adaptive-chosen-ciphertext attack.

During the late 1990s, Daniel Bleichenbacher demonstrated a practical adaptive-chosen-ciphertext attack on SSL servers using a form of the RSA encryption [35]. Almost at the same time, The Cramer-Shoup asymmetric encryption algorithm which is extremely malleable, and an extension of the Elgamal algorithm was proposed [36]. It is the first efficient scheme proven to be secure against the adaptive-chosen-ciphertext attack using standard cryptographic assumptions, which implies that not all uses of cryptographic hash functions require random oracles — some require only the property of collision resistance.

#### 5.3.1 Different Ciphertexts of the Identical Plaintext

It is lucky that REESSE1+ can avoid the adaptive-chosen-ciphertext attack when a REESSE1+ ciphertext is produced according to the following algorithm:

Paralleling Section 3.2, assume that $b_1 \ldots b_n$ is a plaintext block, and $\{C_1, \ldots, C_n\}$ is a public key.

S1: Set $k \leftarrow 0$, $i \leftarrow 1$.

S2: If $b_i = 0$, let $k \leftarrow k + 1$, $b_i \leftarrow 0$;
   else let $b_i \leftarrow k + 1$, $k \leftarrow 0$.

S3: Let $i \leftarrow i + 1$.
   If $i \le n$, goto S2.

S4: Randomly produce $s_1 \ldots s_n \in \{0, 1\}^n$.

S5: If $b_n = 0$, set $d \leftarrow n - k$, $s_d = 1$, $b_d \leftarrow b_d + k$;
   else $s_n = 1$.

S6: Compute $\bar{G} \leftarrow \prod_{i=1}^{n} (C_{i\, s_i + (i - \underset{.}{b}_i + 1) \neg s_i})^{\underset{.}{b}_i} \% M$ with $C_{n+1} = 1$.

Clearly, when the identical plaintext is inputted many times, the above algorithm will return a ciphertext different from one another every time.

It is easily understood that contrarily, a ciphertext can almost uniquely be decrypted in polynomial time in terms of Section 3.3 and 3.9.

Paralleling Section 3.3, we design a related decryption algorithm of which the running time is equivalent to that of the algorithm in Section 3.3.

Assume that ($\{A_i\}$, $\{\ell(i)\}$, $W$, $\delta$, $Đ$, $đ$, $\hbar$) is a related private key, and $\bar{G}$ is a ciphertext.

Notice that because $\sum_{i=1}^{n} \underset{.}{b}_i = n$ is even, $\sum_{i=1}^{n} \underset{.}{b}_i \ell(i)$ must be even.

S1: Compute $\bar{G} \leftarrow \bar{G}^{\delta^{-1}} \% M$.

S2: Compute $\bar{G} \leftarrow \bar{G} W^{-2} \% M$.

S3: Set $b_1 \ldots b_n$, $\ddot{e}, j, k \leftarrow 0$, $G \leftarrow \bar{G}$, $i \leftarrow 1$.

S4: If $A_i^{\ddot{e}+1} \mid G$, let $\ddot{e} \leftarrow \ddot{e} + 1$, goto S4.

S5: If $\ddot{e} = 0$,

  let $k \leftarrow k + 1$, $i \leftarrow i + 1$;

 else

  compute $G \leftarrow G / A_i^{\ddot{e}}$;

  if $k > 0$ or $i + \ddot{e} - 1 = n$, let $b_i \leftarrow 1$; else $b_{i+\ddot{e}-1} \leftarrow 1$;

  if $k = 0$, let $i \leftarrow i + \ddot{e}$; else $i \leftarrow i + 1$;

  if $k + 1 > \ddot{e}$, let $i \leftarrow n + 1$;

  set $\ddot{e}, k \leftarrow 0$.

S6: If $i \le n$ and $G \ne 1$, goto S4.

S7: If $G \ne 1$, goto S2; else end.

In this way, the original plaintext block $b_1 \ldots b_n$ is recovered although $s_1 \ldots s_n$ is introduced in encryption. Besides, in decryption, $\{\ell(i)\}$, $Đ$, $đ$, and $\hbar$ are unhelpful.

### 5.3.2 Appending of a Stochastic Binary String

Another approach to avoiding the adaptive-chosen-ciphertext attack is to append a stochastic fixed-length binary sequence to the terminal of every plaintext block when it is encrypted. For a concrete implementation, refer to the OAEP+ scheme [37].

# 6 Security Analysis of the Digital Signature

The security of the REESSE1+ signature includes three aspects: a private key cannot inferred from a signature, a signature cannot be forged through known signatures, public keys, and algorithms, and a message cannot be forged through a known or chosen signature.

## 6.1 Unforgeability of a Signature in the Random Oracle Model

Because REESSE1+ is a multiproblem cryptosystem, and MPP preventing a forged signature from being obtained from the signature algorithm is different from TLP or PRFP preventing a forged signature from being obtained from the verification algorithm, the proof in the random oracle model (RO model) given below is incomplete, and only offers another piece of evidence for the security of the REESSE1+ signature.

### 6.1.1 What Is a Random Oracle

An oracle is a mathematical abstraction, a theoretical black box, or a subroutine of which the running time may not be considered [25][38]. In particular, in cryptography, an oracle may be treated as a subcomponent of an adversary, and lives its own life independent of the adversary. Usually, the adversary interacts with the oracle but cannot control its behavior.

A random oracle is an oracle which answers to every query with a completely random and unpredictable value chosen uniformly from its output domain, except that for any specific query, it outputs the same value every time it receives that query if it is supposed to simulate a deterministic function [38].

Random oracles are utilized in cryptographic proofs for relpacing any irrealizable function so far which can provide the mathematical properties required by the proof. A cryprosystem or a protocol that is proven secure using such a proof is described as being secure in the RO model, as opposed to being secure in the standard model where IFP, DLP etc are assumed to be hard. When a random oracle is used within a security proof, it is made available to all participants, including adversaries. In practice, random oracles producing a bit-string of infinite length which can be truncated to the length desired are typically used to model cryptographic hash functions in schemes where strong randomness assumptions of a hash function's output are needed.

In fact, it draws attention that certain artificial signature and encryption schemes are proven secure in the RO model, but are trivially insecure when any real function such as the hash function MD5 or SHA-1 is substituted for the random oracle [39]. Nevertheless, for any more natural protocol, a proof of security in the RO model gives very strong evidence that an attacker have to discover some unknown and undesirable property of the hash function used in the protocol.

### 6.1.2 The Forking Lemma

***Lemma 1 (the forking lemma):*** Let $\hat{A}$ be a probabilistic polynomial time Turing machime, given only the public data as input. If $\hat{A}$ can find, with non-negligible probability, a valid signature ($\underline{m}$, $\sigma_1$, $\underset{.}{h}$, $\sigma_2$), then, with non-negligible probability, a replay of this machine, with the same random tape and a different oracle, outputs two valid signatures ($\underline{m}$, $\sigma_1$, $\underset{.}{h}$, $\sigma_2$) and ($\underline{m}$, $\sigma_1$, $\underset{.}{h}'$, $\sigma_2'$) such that $\underset{.}{h} \ne \underset{.}{h}'$ [40].

In [40], the forking lemma is specified in terms of an adversary that attacks a digital signature scheme instantiated in the RO model. D. Pointcheval and J. Stern show that if an adversary can forge a signature with non-negligible probability, then there is a non-negligible probability that the same adversary with the same random tape can create a second forgery in an attack with a different random oracle. The forking lemma was later generalized by M. Bellare and G. Neven [41], and has been used to prove the security of a variety of digital signature schemes and other cryptographic constructions based on random oracles [42].

The forking lemma is applicable to such a type of signature scheme where a signer must perform the following steps to sign the message $\underline{m}$:

① randomly produce a promise $\sigma_1$ in a large set;
② compute $\underline{h} = hash(\sigma_1, \underline{m})$;
③ compute $\sigma_2$ by using $\sigma_1$ and $\underline{h}$.

The signature output is a triple $(\sigma_1, \underline{h}, \sigma_2)$.

For example, the Schnorr signature scheme [43] may be proven secure in the RO model according to the forking lemma [40][42].

Assume that solving $y \equiv g^x$ (% $M$) for $x$ is of DLP, where $g$ is a generator of $\mathbb{Z}_M^*$ with $M$ prime, and the input to an adversary $\hat{A}$ is the public key $y$.

In terms of the forking lemma, the adversary can obtain two different signatures with non-negligible probability:

$$(\underline{m}, \sigma_1 = g^r \,\%\, M, \underline{h}, \sigma_2 = (x\underline{h} + r) \,\%\, \bar{M}), \text{ and } (\underline{m}, \sigma_1' = g^{r'} \,\%\, M, \underline{h}', \sigma_2' = (x\underline{h}' + r') \,\%\, \bar{M}),$$

where $\underline{h} = hash(\sigma_1, \underline{m})$ is an oracle query of the adversary $\hat{A}$ during his the first play, and $\underline{h}' = hash'(\sigma_1', \underline{m})$ is an oracle query of the adversary $\hat{A}$ during his the second play.

The target of the adversary $\hat{A}$ is to extract the private key $x$ form the signatures.

Since there is $\sigma_1 = \sigma_1'$, there is $r = r'$. Hence, we have

$$(\underline{h} - \underline{h}')x \equiv (\sigma_2 - \sigma_2') \ (\%\ \bar{M}).$$

Because there always exists $x$, the above linear congruence in $x$ has solutions, and $x$ can be found in polynomial time, namely DLP can be solved in polynomial time, which is in direct contradiction to the standard assumption. Therefore, it is infeasible to forge a Schnorr signature.

### 6.1.3 Proof of Unforgeability of a REESSE1+ Signature

In Section 4.1, we prove that MPP is computationally at least equivalent to DLP, and additionally at present, a subexponential time algorithm for solving MPP is not found. Therefore, we may have a superstandard assumption comparable to the standard assumption.

***Assumption 1:*** MPP cannot be solved in subexponential time.

Firstly, we adapt the REESSE1+ signature algorithm to the modality outputting a triple signature.

Assume that $(\{A_i\}, \{\ell(i)\}, W, \delta, Đ, đ, \hbar)$is a private, and $F$ is a file or message to be signed.

Define $hash'(F, \bar{a})$ as $hash'(F, \bar{a}) = (\bar{a}Đ + hash(F)W)\delta^{-1}$ % $\bar{M}$.

S1: Let $H \leftarrow hash(F)$, whose binary form is $b_1 \ldots b_n$.

S2: Randomly take $\bar{a} \in (1, \bar{M})$, compute $Q = hash'(F, \bar{a})$. If $(đT) \mid \bar{a}$ or $đ \mid (WQ)$ % $\bar{M}$, goto S2.

S3: Set $\underline{k} \leftarrow \delta \sum_{i=1}^n b_i \ell(i)$ % $\bar{M}$, $G_0 \leftarrow (\prod_{i=1}^n A_i^{\neg b_i})^\delta$ % $M$.

S4: Compute $R \leftarrow (Q(\delta\hbar)^{-1})^{S^{-1}} G_0^{-1}$, $\bar{U} \leftarrow (RW^{\underline{k}-\delta})^Q$ % $M$, $\bar{g} \leftarrow \delta^{\bar{a}Đ}$ % $M$, $\xi \leftarrow \sum_{i=0}^{n-1} (\delta Q)^{n-1-i} (HW)^i$ % $\bar{M}$.

S5: $\forall r \in [1, đ2^{16}]$ making $đ \nmid (rUS + \xi)$ % $\bar{M}$, where $U = \bar{U}\bar{g}^r$ % $M$.

S6: If $d \nmid ((WQ)^{n-1} + rUS + \xi)$ % $\bar{M}$, goto S6, else end.

On executing this algorithm which is essentially equivalent to that in Section 3.4, one can obtain the signature $(\bar{a}, Q, U)$ that is sent to a receiver together with $F$.

Secondly, prove the unforgeability of a signature in the RO model.

*Proof:*

Let $\bar{a}$ be a promise produced by the signer. In terms of the forking lemma, during the two plays of the adversary $\hat{A}$, two valid signatures $(\bar{a}, Q, U)$ and $(\bar{a}', Q', U')$ on the identical file $F$ can be obtained. Then, it is known from the above signature algorithm that

$$\delta Q \equiv \bar{a}Đ + WH \ (\%\ \bar{M}),$$
$$\delta Q' \equiv \bar{a}'Đ + WH' \ (\%\ \bar{M}),$$
$$U \equiv (RW^{\underline{k}-\delta})^Q \delta^{\bar{a}Đr} \ (\%\ M),$$
$$U' \equiv (R'W^{\underline{k}'-\delta})^{Q'} \delta^{\bar{a}'Đr'} \ (\%\ M),$$

where $\bar{a} = \bar{a}'$, $H = hash(F, \bar{a})$, and $H' = hash'(F, \bar{a}')$.

Because there always exist $W$, it may be obtained from the two formulas with $H$ and $H'$ that

$$W \equiv (Q - Q')(H - H')^{-1}\delta \ (\%\ \bar{M}). \quad (2)$$

Similarly, it may be obtained from the two formulas with $U$ and $U'$ that

$$(R^Q R'^{-Q'})(W^{(\underline{k}-\delta)Q} W^{-(\underline{k}'-\delta)Q'}) \delta^{-\bar{a}Đ(r-r')} \equiv U U'^{-1} \ (\%\ M).$$

Multiplying either side of the above equation by $G_1^Q G_1'^{-Q'}$ gives

$$(R^Q R'^{-Q'})(W^{(\underline{k}-\delta)Q} G_1^Q W^{-(\underline{k}'-\delta)Q'} G_1'^{-Q'}) \delta^{-\bar{a}Đ(r-r')}$$
$$\equiv U G_1^Q U'^{-1} G_1'^{-Q'} \ (\%\ M),$$

namely

$$((Q(\delta\hbar)^{-1})^{S^{-1}} G_0^{-1})^Q ((Q'(\delta\hbar)^{-1})^{S^{-1}} G_0'^{-1})^{-Q'} (\bar{G}_1^Q W^{-\delta Q} \bar{G}_1'^{-Q'} W^{\delta Q'}) \delta^{\bar{a}Đ(r-r')}$$
$$\equiv U G_1^Q U'^{-1} G_1'^{-Q'} \ (\%\ M),$$

where $G_0' = (\prod_{i=1}^n A_i^{\neg b_i'})^\delta$ (% $M$) and $G_1' = (\prod_{i=1}^n A_i^{b_i'})^\delta$ (% $M$).

Further, there is

$$(\delta\hbar)^{-S^{-1}(Q-Q')} Q^{QS^{-1}} Q'^{-Q'S^{-1}} (G_0^{-1}\bar{G}_1)^Q (G_0'^{-1}\bar{G}_1')^{-Q'} (W^{-\delta})^{Q-Q'} \delta^{-\bar{a}Đ(r-r')}$$
$$\equiv U G_1^Q U'^{-1} G_1'^{-Q'} \ (\%\ M).$$

Through transposition, we see

$$((\delta\hbar)^{-S^{-1}} W^{-\delta})^{Q-Q'} \delta^{-\bar{a}Đ(r-r')}$$
$$\equiv U G_1^Q U'^{-1} G_1'^{-Q'} Q^{-QS^{-1}} Q'^{Q'S^{-1}} (G_0^{-1}\bar{G}_1)^{-Q} (G_0'^{-1}\bar{G}_1')^{Q'} \ (\%\ M).$$

Substituting $W^{-\delta}$ with $(\hbar\delta\alpha^{-1})^{S^{-1}} (\prod_{i=1}^n A_i)^\delta \equiv (\hbar\delta\alpha^{-1})^{S^{-1}} G_0 G_1$ (% $M$) yields

$$((\delta\hbar)^{-S^{-1}} (\hbar\delta\alpha^{-1})^{S^{-1}} G_0 G_1)^{Q-Q'} \delta^{-\bar{a}Đ(r-r')}$$
$$\equiv (G_0G_1)^Q (G_0'G_1')^{-Q'} U U'^{-1} Q^{-QS^{-1}} Q'^{Q'S^{-1}} \bar{G}_1^{-Q} \bar{G}_1'^{Q'} \ (\%\ M),$$

namely

$$(\alpha^{-S^{-1}} G_0 G_1)^{Q-Q'} \delta^{-\bar{a}Đ(r-r')}$$
$$\equiv (G_0G_1)^Q (G_0'G_1')^{-Q'} U U'^{-1} Q^{-QS^{-1}} Q'^{Q'S^{-1}} \bar{G}_1^{-Q} \bar{G}_1'^{Q'} \ (\%\ M).$$

Again through transposition, we see

$$\delta^{-\bar{a}Đ(r-r')} \equiv$$
$$(G_0G_1)^Q (G_0'G_1')^{-Q'} U U'^{-1} Q^{-QS^{-1}} Q'^{Q'S^{-1}} \bar{G}_1^{-Q} \bar{G}_1'^{Q'} (\alpha^{-S^{-1}} G_0 G_1)^{Q'-Q} (\% M).$$

Considering $G_0G_1 \equiv G_0'G_1' \equiv (\prod_{i=1}^n A_i)^\delta$ (% $M$), we have

$$\delta^{-\bar{a}Đ(r-r')} \equiv U U'^{-1} Q^{-QS^{-1}} Q'^{Q'S^{-1}} \bar{G}_1^{-Q} \bar{G}_1'^{Q'} \alpha^{S^{-1}(Q-Q')} (\% M). \quad (3)$$

In (3), the right value is known, the promise $\bar{a}$ is also known, $D$, a factor of $\bar{M}$, may be separated in subexponential time, and $r, r' \in [1, đ2^{16}]$, $(r - r')$ may be found in subexponential time through a heuristic approach, which means the integer $\bar{a}Đ(r - r')$ may be obtained in subexponential time.

Solving (3) for $\delta$ is of RFP of which the complexity is not greater than that of DLP. Thus, $\delta$ may be found in subexponential time, and then according to (2), $W$ may be found.

In addition, by using $C_i \equiv (A_i W^{\ell(i)})^\delta$ (% $M$), $A_i \in \{2, 3, \ldots, 1201\}$, $\ell(i) \in \{5, 7, \ldots, 2n + 1\}$, and $\{A_1, \ldots, A_n\}$ being a coprime sequence, the adversary may obtain $\{A_1, \ldots, A_n\}$ and $\{\ell(1), \ldots, \ell(n)\}$ in expected subexponential time.

Further, according to $\bar{U} \equiv (RW^{\underline{k}-\delta})^Q$ (% $M$) in the signature algorithm, where $\underline{k} \equiv \delta \sum_{i=1}^n b_i \ell(i)$ (% $\bar{M}$), $R$ may be found, and according to $R \equiv (Q(\delta\hbar)^{-1})^{S^{-1}} G_0^{-1}$ (% $M$), $\hbar$ may be found. In sum, the private key ($\{A_i\}$, $\{\ell(i)\}$, $W$, $\delta$, $Đ$, $đ$, $\hbar$) is found, namely MPP is also solved in subexponential time, which is in direct contradiction to assumption 1. Therefore, it is infeasible in subexponential time that the adversary forge a REESSE1+ promise signature ($\bar{a}$, $Q$, $U$).

In practice, the promise $\bar{a}$ is unpublicized, and therefore, it is at least the same infeasible in subexponential time that $\hat{A}$ forges a REESSE1+ signature ($Q$, $U$). □

Notice that the REESSE1+ verification algorithm is built on the hardnesses different from MPP, so this proof is incomplete for the unforgeability of a REESSE1+ signature, and further security analysis is necessary.

## 6.2 Extracting a Private Key from a Signature Is of TLP or the Indeterminate Problem

To analyze exact securities, we attend to the solution of $x^k \equiv c$ (% $M$) with $M$ prime called RFP for a while. In some cases, $x^k \equiv c$ (% $M$) has the trivial root which can be found in terms of Theorem 1 or 2. At present, there should be three methods of solving $x^k \equiv c$ (% $M$): ① the algorithm through discrete logarithms [25] whose running time is $L_M[1/3, 1.923]$; ② the probabilistic algorithm [44] whose running time is the larger of $O(2^{k-1})$ and $O(M/k)$; ③ the algorithm offered by [45] whose running time is $O(k^{1/2} \lg M)$. Obviously, when $k$ is comparatively small, method 3 is most efficient.

In our analysis, we assume that RFP can be solved in tolerable time.

It is known from Section 3.4 that there exist

$$Q \equiv (RG_0)^S \delta\hbar \ (\% M),$$
$$U \equiv (RW^{ķ-\delta})^Q \delta^{\bar{a}Đr} \ (\% M).$$

Firstly, solving the two congruences separately.

According to Section 3.4, $\delta^{\bar{a}Đr}$ % $M$ belongs to the subgroup of order $đT$. Because of $T \ge 2^n$, divining the value of $\delta^{\bar{a}Đr}$ % $M$ is impossible.

Let $\bar{e} = \delta^{\bar{a}ĐrT}$ % $M$ be an element of the subgroup of order $đ$, then latter may be written as

$$U^T \equiv (RW^{ķ-\delta})^{QT} \bar{e} \ (\% M).$$

When an attacker attempts to seek $RG_0$ or $RW^{ķ-\delta}$, he has to solve the equation

$$x^S \equiv Q\delta^{-1}\hbar^{-1} \ (\% M), \quad (4)$$

or

$$x^{QT} \equiv U^T \bar{e}^{-1} \ (\% M). \quad (5)$$

We see that the number of unknown variables in the two equations is greater than 2, and thus solving (4) and (5) is of the indeterminate problem.

For (4), because $\delta$, $\hbar$ are unknown, and the right of (4) is not a constant, it is impossible to solve this equation for $RG_0$. If $\delta$ is divined, the probability of hitting $\delta$ is $1/\|\delta\| < 1/2^n$. Similarly it is impossible to divine $\hbar$ owing to $\hbar \in (1, M)$.

For (5), there is $\|\bar{e}^{-1}\| \le đ$. Assume that $đ$ is guessed out, and all the solutions to $x^{đ} \equiv 1$ (% $M$) can be found out, then $\bar{e}^{-1}$ may possibly be figured.

However, even if $\bar{e}^{-1}$ is known, and the trivial root to (5) exists, the probability that the trivial root just equals the specific $RW^{ķ-\delta}$ is only $1/T \le 1/2^n$, and moreover due to the randomicity of $R$, it is thoroughly impossible to separate $\delta$, $W$, and $ķ$ from $RW^{ķ-\delta}$.

Secondly, solving the two congruences systematically.

Substituting $R$ in $U^T \equiv (RW^{ķ-\delta})^{QT} \bar{e}$ (% $M$) with $G_0^{-1}(Q(\delta\hbar)^{-1})^{S^{-1}}$ derived from $Q \equiv (RG_0)^S \delta\hbar$ (% $M$) gives

$$U^T \equiv (G_0^{-1}(Q(\delta\hbar)^{-1})^{S^{-1}} W^{ķ-\delta})^{QT} \bar{e} \ (\% M),$$

namely

$$U^T \equiv (G^{-1}\bar{G}_1(Q(\delta\hbar)^{-1})^{S^{-1}} W^{-\delta})^{QT} \bar{e} \ (\% M),$$

where $G \equiv G_0 G_1$ (% $M$). Since $W$ is a function of $\delta$ in $\beta = \delta^{W^n T}$ % $M$, $y = W^{-\delta}$ % $M$ is of TLP.

Thus,

$$((GW^{\delta})^{-1}(\delta\hbar)^{-S^{-1}})^{QT} \equiv U^T(\bar{G}_1 Q^{S^{-1}})^{-QT} \bar{e}^{-1} \ (\% M).$$

Similarly, if $\bar{e}^{-1}$ is known, and $(U^T(\bar{G}_1 Q^{S^{-1}})^{-QT} \bar{e}^{-1})^{\bar{M}/(T\gcd(Q,\bar{M}))} \equiv 1$ (% $M$), through the Index-calculus method, one may find out all the solutions to the equation

$$x^{QT} \equiv U^T(\bar{G}_1 Q^{S^{-1}})^{-QT} \bar{e}^{-1} \ (\% M).$$

However, the probability that a certain solution is no other than $(GW^{\delta})^{-1}(\delta\hbar)^{-S^{-1}}$ is less than $1/T \le 1/2^n$. Further, the running time of distinguishing $G$, $W$, $\hbar$, and $\delta$ from $(GW^{\delta})^{-1}(\delta\hbar)^{-S^{-1}}$ is at least $O(\bar{M})$.

Therefore, the time complexity of extracting a related private key from a signature is at least $O(\|\delta\|)$, $O(T)$, or $O(\bar{M}) \ge O(2^n)$, which elucidates that even if each element of the subgroup of order $đ$ is known, it does not influence the security of the private key.

## 6.3 Forging a Digital Signature only from a Public Key Is a Hardness

According to Section 3.5, the discriminant $X \equiv Y$ (% $M$) contains the two variables $Q$ and $U$ of which one may be supposed in advance by an adversary. However, seeking the other by the supposed value is faced with TLP.

### 6.3.1 TLP Is at Least Equivalent to DLP

We observe $y \equiv g^x$ (% $M$) referred to as TLP by Definition 7.

Assume that $g \in \mathbb{Z}_M^*$ is a generator, where $M$ is prime, then

$$\{y \mid y \equiv g^x \ (\% M), x = 1, \ldots, \bar{M}\} = \mathbb{Z}_M^* \ [22].$$

Assume that $k$ with $\gcd(k, \bar{M}) = 1$ is an integer, then also

$$\{y \mid y \equiv x^k \ (\% M), x = 1, \ldots, \bar{M}\} = \mathbb{Z}_M^* \ [22].$$

Therefore, $\forall x \in [1, \bar{M}]$, $y \equiv g^x$ (% $M$) or $y \equiv x^k$ (% $M$) with $\gcd(k, \bar{M}) = 1$ is a self-isomorph of the group $\mathbb{Z}_M^*$.

However, for the $x^x$ operation, $\{y \mid y \equiv x^x \ (\% M), x = 1, \ldots, \bar{M}\} = \mathbb{Z}_M^*$ does not hold, that is,

$$\{y \mid y \equiv x^x \ (\% M), x = 1, \ldots, \bar{M}\} \ne \mathbb{Z}_M^*.$$

For example, when $M = 11$, $\{y \mid y \equiv x^x \ (\% M), x = 1, \ldots, \bar{M}\} = \{1, 3, 4, 5, 6\}$, where $3^3 \equiv 6^6 \equiv 8^8 \equiv 5$ (% 11).

When $M = 13$, $\{y \mid y \equiv x^x \ (\% M), x = 1, \ldots, \bar{M}\} = \{1, 3, 4, 5, 6, 9, 12\}$, where $7^7 \equiv 11^{11} \equiv 6$ (% 13), and $1^1 \equiv 3^3 \equiv 8^8 \equiv 9^9 \equiv 12^{12} \equiv 1$ (% 13).

When $M = 17$, $\{y \mid y \equiv x^x \ (\% M), x = 1, \ldots, \bar{M}\} = \{1, 2, 4, 8, 9, 10, 12, 13, 14\}$, where $2^2 \equiv 12^{12} \equiv 4$ (% 17), $6^6 \equiv 15^{15} \equiv 2$ (% 17), and $10^{10} \equiv 14^{14} \equiv 2$ (% 17).

The above examples illustrate that $\{y \equiv x^x \ (\% M) \mid x = 1, \ldots, \bar{M}\}$ cannot construct a complete set for a group. Furthermore, mapping from $x$ to $y$ is one-to-one sometimes, and many-to-one sometimes. That is, inferring $x$ from $y$ is indeterminate, $x$ is nonunique, and even inexistent. Thus, $x^x$ has extremely strong irregularity, and is essentially distinct from $g^x$ and $x^k$.

It should be noted that an attempt at solving $y \equiv x^x$ (% $M$)

for $x$ in light of the Chinese Remainder Theorem is specious. Refer to the following example.

Observe the congruent equation $4^4 \equiv 8 \equiv 3^{12}$ (% 31), where $3 \in \mathbb{Z}_{31}^*$ is a generator.

Try to seek $x$ which satisfies $x \equiv 12$ (% 30) and $x \equiv 3$ (% 31) at one time, and verify whether $x \equiv 4$ (% 31) or not.

In light of the Chinese Remainder Theorem [25], let $m_1 = 30$, $m_2 = 31$, $a_1 = 12$, and $a_2 = 3$. Then

$$M = 30 \times 31 = 930,$$
$$M_1 = M / m_1 = 930 / 30 = 31,$$
$$M_2 = M / m_2 = 930 / 31 = 30.$$

Compute $y_1 = 1$ such that $M_1 y_1 \equiv 1$ (% $m_1$).

Compute $y_2 = 30$ such that $M_2 y_2 \equiv 1$ (% $m_2$).

Thereby,

$$x = a_1 M_1 y_1 + a_2 M_2 y_2$$
$$= 12 \times 31 \times 1 + 3 \times 30 \times 30 = 282 \text{ (\% 930)}.$$

It is not difficult to verify

$$282^{288} \equiv 8 \neq 4 \text{ (\% 31), and } 282^{288} \equiv 504 \neq 8 \neq 4 \text{ (\% 930)}.$$

The integer 282 is an element of the group $(\mathbb{Z}_{930}^*, \cdot)$, and the element 4 of the group $(\mathbb{Z}_{31}^*, \cdot)$ cannot be obtained from 282, which is pivotal.

So, these examples manifest that TLP seems to be harder than DLP.

***Property 7:*** TLP is computationally at least equivalent to DLP in the same prime field.

*Proof:*

① $\hat{H}(y \equiv x^x \text{ (\% } M)) =_T^P \hat{H}(y \equiv (gx)^x \text{ (\% } M))$

Let $g \in \mathbb{Z}_M^*$ be a generator coprime to $\bar{M}$, which does not lose generality since $g$ may be selected in practice.

Assume that $y \in \mathbb{Z}_M^*$ is known, and there is

$$y \equiv (gx)^x \text{ (\% } M).$$

Raising either side of the above equation to the $g$-th power gives

$$y^g \equiv (gx)^{gx} \text{ (\% } M).$$

Let

$$z \equiv y^g \text{ (\% } M), \text{ and } w = gx,$$

where the latter is not a congruence, and then

$$z \equiv w^w \text{ (\% } M).$$

Suppose that $\bar{O}_t(y, M, X)$ is an oracle on solving $y \equiv x^x$ (% $M$) for $x$, where $X$ is the set of all the possible values of $x$, and $y \in [1, \bar{M}]$.

Its output is $x \in X$ (each of solutions), or 0 (no solution).

Let $X_1 = \{1, 2, \ldots, \bar{M}\}$, and $X_2 = \{1g, 2g, \ldots, \bar{M}g\}$.

Clearly, by calling $\bar{O}_t(y, M, X_1)$, $y \equiv x^x$ (% $M$) is solved for $x$.

It is easily observed that between the finite sets $X_1$ and $X_2$, there is a linear bijection

$$\Gamma: X_1 \to X_2, \ \Gamma(a) = ga,$$

which means that the set $X_1$ is equivalent to the set $X_2$ [46]. Hence, substituting $X_1$ with $X_2$ as the codomain of a function will not increase the running time of $\bar{O}_t$.

Similarly, by calling $\bar{O}_t(z, M, X_2)$, $z \equiv w^w$ (% $M$) is solved for $w$, namely all the satisfactory values of $w$ are obtained.

According to $w = gx$ and $z \equiv w^w$ (% $M$), there is $x \equiv wg^{-1}$ (% $M$), or $x \equiv wg^{-1}$ (% $\bar{M}$).

Hence, in terms of Definition 8, there is

$$\hat{H}(y \equiv (gx)^x \text{ (\% } M)) \le_T^P \hat{H}(y \equiv x^x \text{ (\% } M)),$$

namely the difficulty in inverting $y \equiv (g\,x)^x$ (% $M$) is not greater than that in inverting $y \equiv x^x$ (% $M$).

On the other hand, suppose that $\bar{O}_t(y, g, M)$ is an oracle on solving $y \equiv (gx)^x$ (% $M$) for $x$, where $y, g \in [1, \bar{M}]$.

Its output is $x \in [1, \bar{M}]$ (each of solutions), or 0 (no solution).

Let $g = 1$.

By calling $\bar{O}_t(y, 1, M)$, the solution $x$ to $y \equiv x^x$ (% $M$) will be obtained.

Hence, in terms of Definition 8, there is

$$\hat{H}(y \equiv x^x \text{ (\% } M)) \le_T^P \hat{H}(y \equiv (gx)^x \text{ (\% } M)).$$

Combinatorially, in terms of Definition 9, we have that

$$\hat{H}(y \equiv x^x \text{ (\% } M)) =_T^P \hat{H}(y \equiv (gx)^x \text{ (\% } M)),$$

namely the difficulty in inverting $y \equiv (g\,x)^x$ (% $M$) is equivalent to that in inverting $y \equiv x^x$ (% $M$).

② $\hat{H}(y \equiv g^x \text{ (\% } M)) \le_T^P \hat{H}(y \equiv (gx)^x \text{ (\% } M))$

The congruence $y \equiv (gx)^x$ (% $M$) may be written as $y \equiv g^x x^x$ (% $M$), where $g$ is any element.

Change $\bar{O}_t(y, g, M)$ into $\bar{O}_t(y, g, M, \hat{w})$, where $\hat{w} = 0$ or 1. Its structure is as follows:

S1: If $\hat{w} = 1$ and $x$ to $y \equiv g^x x^x$ (% $M$) inexistent, return 'No'.

S2: If $\hat{w} = 1$,

S2.1: find $y_1$, and compute $y_2$ by $y \equiv y_1 y_2$ (% $M$);

S2.2: compute $x < M$ by $y_1 \equiv g^x$ (% $M$);

S2.3: if $y_2 \neq x^x$ (% $M$), goto S2.1;

else

S2.4: compute $x < M$ by $y \equiv g^x$ (% $M$).

S3: Return $x$.

Clearly, by calling $\bar{O}_t(y, g, M, 0)$, the solution $x$ to $y \equiv g^x$ (% $M$) will be obtained.

Hence, still in terms of Definition 8, there is

$$\hat{H}(y \equiv g^x \text{ (\% } M)) \le_T^P \hat{H}(y \equiv g^x x^x \text{ (\% } M)).$$

Integrating ① and ②, we have that

$$\hat{H}(y \equiv g^x \text{ (\% } M)) \le_T^P \hat{H}(y \equiv g^x x^x \text{ (\% } M)) =_T^P \hat{H}(y \equiv x^x \text{ (\% } M)),$$

namely inverting $y \equiv x^x$ (% $M$) is at least equivalent inverting $y \equiv g^x$ (% $M$) for $x$ in complexity. □

In [47], we have a similar result by the asymptotic granularity reduction (AGR).

Further discussion.

Let $y \equiv g^v$ (% $M$), and $x \equiv g^u$ (% $M$), and then it seems that there is $g^v \equiv g^{ug^u}$ (% $M$).

However due to $g^u$ (% $M$) $\neq g^u$ (% $\bar{M}$), $y \equiv x^x$ (% $M$) cannot be expressed as $v \equiv ug^u$ (% $\bar{M}$).

We can also understand that in the process of $x$ being sought from $y \equiv x^x$ (% $M$), it is inevitable that the middle value of $x$ is beyond $M$ because modular multiplication, inverse, and power operations are inevitable.

Considering the middle value of $x$ beyond $M$, let

$$z_1 = x \text{ \% } M \text{ with } z_1 < M, \text{ and } z_2 = x \text{ \% } \bar{M} \text{ with } z_2 < \bar{M}.$$

Then there are $x = z_1 + k_1 M = z_2 + k_2 \bar{M}$ and $z_1 = (z_2 - k_2)$ % $M$, where $k_1$, $k_2 \ge 0$ are two integers. Ahead, we have $y \equiv (g(z_2 - k_2))^{z_2}$ (% $M$). This formula indicates that due to $x$ % $M \neq x$ % $\bar{M}$ with $x > M$ and $k_2$ unfixable, the relation between $x$ % $\bar{M}$ and $x$ % $M$ is stochastic when $x$ changes in the interval $(1, M^M)$, which illuminates that it is reasonable to let $q \equiv g(z_2 - k_2)$ (% $M$), namely $y \equiv q^{z_2}$ (% $M$).

If $q$ is a constant, inverting $y \equiv q^{z_2}$ (% $M$) is equivalent to DLP. However, $q$ will not be a constant forever. Therefore, it should be impossible anyway that $\hat{H}(y \equiv (gx)^x \text{ (\% } M)) =_T^P \hat{H}(y \equiv g^x \text{ (\% } M))$.

The above evidence inclines us to believe that TLP is harder than DLP, namely on the assumption that DLP can be solved through an oracle, TLP cannot be solved in DLP subexponential time yet.

The famous baby-step giant-step algorithm, Pollard's rho algorithm, Pohlig-hellman algorithm, and index-calculus algorithm for discrete logarithms [25] are ineffectual on transcendental logarithms. At present, there is no better method for solving TLP than exhaustive search, and thus the running time of solving $x^x \equiv y$ (% $M$) may be expected to be $O(M) > O(2^n)$.

Notice that for $y \equiv x^x$ (% $M$), there is no determinate relation between $\|y\|$ and $\|x\|$, namely $\|y\| \geq \|x\|$ or $\|y\| < \|x\|$. Therefore, in the case of a small modulus, $x$ in $y \equiv x^x$ (% $M$) is still secure.

In REESSE1+, the form of TLP is $y \equiv (cx)^x$ (% $M$) with $c$ known. When the bit-length of a modulus is very small — 80 for example, the difference between the running times of solving $y \equiv (cx)^x$ (% $M$) and $y \equiv x^x$ (% $M$) is valuable, where $cx$ changes with $c$, and has more freedom than a single $x$, which makes the relation between $\|y\|$ and $\|x\|$ be more indeterminate.

What needs to be emphasized is that TLP is more suitable for designing signature schemes due to the non-uniqueness of its solutions.

### 6.3.2 Forging a Signature from the Verification Algorithm Is of TLP or PRFP

Assume that $H$ is the output of *hash* on input of a file $F$, and $(Q, U)$ is a signature on $F$. According to the discriminant $X \equiv Y$ (% $M$), namely

$$(\alpha Q^{-1})^{QUT} \alpha^{Q^n} \equiv (\bar{G}_1^{Q} U^{-1})^{UST} \beta^{HQ^{n-1} + H^n} \text{ (\% } M),$$

an adversary may seek the value of any signature variable by supposing the value of the other variable.

If suppose the value of $Q$, no matter whether $U$ exists or not, seeking $U$ is equivalent to TLP.

Similarly, if suppose the value of $U$, seeking $Q$ is also equivalent to TLP.

① Faced with RFP and the tight constraint

If the adversary hits exactly the small $\text{đ}$, raises either side of the discriminant to the $\text{đ}$-th power, and assumes that $\text{Đ} \mid (\delta Q - WH)$ holds, then there is

$$(\alpha Q^{-1})^{\text{đ}QUT} \equiv (\bar{G}_1^{Q} U^{-1})^{\text{đ}UST} \text{ (\% } M).$$

Further, let

$$(\alpha Q^{-1})^{\text{đ}QT} \equiv (\bar{G}_1^{Q} U^{-1})^{\text{đ}ST} \text{ (\% } M). \tag{6}$$

Now, suppose that $Q$ is known, and $U$ is unknown. If the congruence

$$U^{\text{đ}T} \equiv ((\alpha^{-1} Q)^{\text{đ}QT})^{S^{-1}} \bar{G}_1^{Q\text{đ}T} \text{ (\% } M)$$

has the trivial solution, work $U$ out by Theorem 2; otherwise work $U$ out by the Index-calculus method. However, $Q$ and $U$ must satisfy the constraint $\text{Đ} \mid (\delta Q - WH)$. The probability of satisfaction is at most $1 / \text{Đ} < 1 / 2^n$ when $\delta$ and $W$ are unknown. Notice that successive values of $Q$ cannot guarantee the integral succession of $(\delta Q - WH)$, and $(Q, U = ((\alpha^{-1} Q)^Q)^{S^{-1}} \bar{G}^Q$ % $M)$ does not satisfy the discriminant.

② Faced with PRFP

We observe that $(\alpha^{Q^n} \beta^{-(HQ^{n-1} + H^n)})^{\text{đ}} \equiv 1$ (% $M$), namely $\alpha^{Q^n} \beta^{-(HQ^{n-1} + H^n)}$ is an element of the subgroup of order $\text{đ}$.

Assume that $\bar{e}$ is a solution to $x^{\text{đ}} \equiv 1$ (% $M$), and $g$ is a generator of $\mathbb{Z}_M^*$. Evaluate $u$, $v$, $q$ by the Index-calculus such that $g^u \equiv \alpha$ (% $M$), $g^v \equiv \beta$ (% $M$), and $g^q \equiv \bar{e}$ (% $M$) [25]. Then

$$g^{uQ^n} g^{-v(HQ^{n-1} + H^n)} \equiv g^q \text{ (\% } M),$$

namely

$$uQ^n - vHQ^{n-1} - vH^n - q \equiv 0 \text{ (\% } \bar{M}) \tag{7}$$

which is a modular polynomial equation in $Q$.

If this polynomial equation has solutions, and $Q$ can be figured out, $U$ may be evaluated according to (6). In this wise, $Q$ and $U$ which likely meet the discriminant can be found.

Solving (7) for $Q$ is of PRFP for which a generic subexponential algorithm is not found so far, and in terms of AGR [47], PRFP is believed to be harder than RFP. Besides, because $\gcd(u, \bar{M}) > 1$ is fully possible, namely (7) is not necessarily a polynomial of which the first term coefficient is 1, and there exists $\bar{M}^{1/n} \in (2^{696/80}, 2^{1216/128}) \approx (2^{8.7}, 2^{9.5})$, attack on (7) is ineffectual by the Coppersmith reduction which can find sufficiently small solutions, whose absolute values are less than $\bar{M}^{1/n}$, to a modular univariate polynomial [48] if such solutions exist. Again, if the adversary solves (7) through the probabilistic algorithm in Section 1.6 of [44], the time complexity will be $O(\bar{M} / n) > O(2^n)$.

③ Unmalicious subgroup of order $\text{đ}$

We observe from (6) that if $\text{đ}$ is guessed accurately, and $\gcd(U, \bar{M}) = 1$ holds, there is

$$((\alpha Q^{-1})^{QT} (\bar{G}_1^{-Q} U)^{ST})^{\text{đ}} \equiv 1, \text{ or } ((\alpha Q^{-1})^{Q} (\bar{G}_1^{-Q} U)^{S})^{\text{đ}T} \equiv 1 \text{ (\% } M),$$

which implies that the element $(\alpha Q^{-1})^{QT} (\bar{G}_1^{-Q} U)^{ST}$ belongs to the subgroup of order $\text{đ}$. Therefore, if gather many enough signatures $(Q, U)$, all the elements of the subgroup are likely found out. However, the analysis in Section 6.2 shows that even if this case occurs, it does not influence the security of a REESSE1+ signature.

Further, through gathering more enough signatures or following the Index-calculus method, all the elements of the subgroup of order $\text{đ}T$ are likely found out. They can be described with a general expression in the time $L_M[1 / 3, 1.923]$, but picking out a specific element will take the time $O(\text{đ}T) > O(2^n)$ since we must try all the elements one by one.

### 6.3.3 Forging a Signature from the Signature Algorithm Is of TLP or PRFP

Owing to $Q \equiv (RG_0)^S \delta \hbar$ (% $M$), $U^T \equiv (RW^{\underline{k}-\delta})^{QT} \bar{e}$ (% $M$), and $V \equiv (R^{-1} W^{\delta} G_1)^{QU} \delta^{\lambda}$ (% $M$) (see Section 3.4), an adversary may attempt the following attack approach.

Let

$$Q \equiv a^S \delta \hbar \text{ (\% } M), \; U^T \equiv b^{QT} \bar{e} \text{ (\% } M), \; V \equiv c^{QU} \delta^{\lambda} \text{ (\% } M),$$

where $\lambda$ meets

$$\lambda S \equiv ((WQ)^{n-1} + \xi + rUS)(\delta Q - HW) \text{ (\% } \bar{M}).$$

Correspondingly, there are $ac \equiv (\alpha \delta^{-1} \hbar^{-1})^{S^{-1}}$, and $bc \equiv \bar{G}_1$ (% $M$).

If $\bar{e}$ is hit, $(Q, U)$ is a known signature, and $(U^T \bar{e}^{-1})^{\bar{M}/(T \gcd(Q, \bar{M}))} \equiv 1$ (% $M$) holds, a solution $b$ to the equation $b^{QT} \equiv U^T \bar{e}^{-1}$ (% $M$) can be found through either the Index-calculus method or the trivial root method. Further, $c$ can be figured from $bc \equiv \bar{G}_1$ (% $M$). However, it is impossible to find $a$ from $ac \equiv (\alpha \delta^{-1} \hbar^{-1})^{S^{-1}}$ (% $M$) since $\delta$, $\hbar$ cannot be obtained from $Q$, $U$, and $V$.

If $\delta$, $\hbar$ are found, then $c$, $b$ can be evaluated on assuming a value of $a$. Further, if Đ is extracted from $\bar{M}$, $W$ is figured, and $r$ is guessed, the adversary may compute the values of $Q$ and $U$ which satisfy

$\text{Đ} \mid (\delta Q - WH)\ (\%\ \bar{M})$, and $\text{đ} \mid ((WQ)^{n-1} + \xi + rUS)\ (\%\ \bar{M})$.

If $\delta$, $W$ can be found, $\hbar$ may possibly be computed according to $\alpha \equiv \delta\hbar(W^{\delta}G_0G_1)^S\ (\%\ M)$.

The above analysis shows that the acquiring $\delta$ and $W$ is the key to the problem. To seek $\delta$ and $W$ from a known clear clue, the adversary has to try to solve the simultaneous equations

$$\begin{cases} \alpha \equiv \delta^{(\delta^n + \delta W^{n-1})T}\ (\%\ M) \\ \beta \equiv \delta^{W^n T}\ (\%\ M). \end{cases}$$

Obviously, the first equation is computationally at least equivalent to TLP. The second equation contains two variables, and belongs to the indeterminate problem. Raising either side of the first to the $W$-th power yields

$$\alpha^W \equiv \delta^{(\delta^n W + \delta W^n)T} \equiv \delta^{\delta^n WT}\beta^{\delta}\ (\%\ M),$$

which is still very complicated, and the problem is not simplified.

Let $g$ be a generator of the group $(\mathbb{Z}_M^*, \cdot)$. By the Index-calculus for discrete logarithms [25], evaluate $u$, $v$, and $x$ such that $g^u \equiv \alpha$, $g^v \equiv \beta$, and $g^x \equiv \delta\ (\%\ M)$ (notice that the latter does not mean $g^x \equiv \delta\ (\%\ \bar{M})$). Then, we have

$$\begin{cases} u \equiv xT(\delta^n + \delta W^{n-1})\ (\%\ \bar{M}) \\ v \equiv xTW^n\ (\%\ \bar{M}). \end{cases}$$

If $x$ is guessed, the two values of $W$ may possibly be obtained according to the above two equations. Nevertheless the two values of $W$ are not necessarily equal to each other, which indicates that the value of $x$ is not right.

If the inverse of $W\ \%\ \bar{M}$ exists, there is $xT \equiv vW^{-n}\ (\%\ \bar{M})$. Hence, we have

$$\begin{aligned} u &\equiv vW^{-n}(\delta^n + \delta W^{n-1}) \\ &\equiv v((\delta W^{-1})^n + \delta W^{-1})\ (\%\ \bar{M}) \\ &\equiv v(y^n + y)\ (\%\ \bar{M}), \end{aligned}$$

where $y \equiv \delta W^{-1}\ (\%\ \bar{M})$. Notice that $y$ is of great value to the adversary if it can be found, for the adversary may obtain $Q \equiv y^{-1}H\ (\%\ \bar{M})$ as he lets $\delta Q - WH \equiv 0\ (\%\ \bar{M})$ according to $\text{Đ} \mid (\delta Q - WH)$.

The above congruence is of PRFP which is very intractable, and believed to be harder than RFP [47]. Additionally, because $\gcd(v, \bar{M}) > 1$ is fully possible, and there exists $\bar{M}^{1/n} \in (2^{696/80}, 2^{1216/128}) \approx (2^{8.7}, 2^{9.5})$, attack on the above congruence is ineffectual by the Coppersmith reduction [48]. If you are afraid of the Coppersmith reduction, in practice, the exponent $n$ independent of the block length may be substituted with a larger integer.

If $\bar{M}$ is factorized, Đ is obtained, and $W^{-1}\ \%\ \text{Đ}$ exists, there is $u \equiv v(y^n + y)\ (\%\ \text{Đ})$, where Đ contains a prime not less than $2^n$. When the Coppersmith reduction is useless, the adversary may attempt the exhaustive search. Nevertheless, its running time is close to $\text{Đ} \geq 2^n$.

Again it is known from the key generation algorithm that there is $W \equiv (\prod_{i=1}^{n} A_i)^{-1}(\alpha\delta^{-1}\hbar^{-1})^{(S\delta)^{-1}}\ (\%\ M)$. However, such a substitution of $W$ will not make the simultaneous equations reduced since $\prod_{i=1}^{n} A_i$ is unknown, and $W(\prod_{i=1}^{n} A_i)$ is the $(S\delta)^{-1}$-th power of the unknown quantity $(\alpha\delta^{-1}\hbar^{-1})$.

## 6.4 Forging a Signature from Known Signatures with a Public Key Is of TLP or *Hash*$^{-1}$

Given a file $F$ and a signature $(Q, U)$ on it, and assume that there exists another file $F'$ with the related $H'$ and $\bar{G}_1'$. If any arbitrary $(Q', U')$ satisfies

$$(\alpha Q'^{-1})^{Q'U'T}\alpha^{Q'^n} \equiv (\bar{G}_1'^{Q'}U'^{-1})^{U'ST}\beta^{H'Q'^{n-1} + H'^n}\ (\%\ M),$$

it is a signature fraud on $F'$.

Clearly, an adversary is allowed to utilize the known values of $Q$ and $U$ separately.

If let $Q' = Q$, $Q'$ does not necessarily satisfy $\text{Đ} \mid (\delta Q' - WH')$, and computing $U'$ is equivalent to TLP.

If let $U' = U$, no matter whether the discriminant has solutions or not, seeking $Q'$ is also equivalent to TLP.

If the two signatures $(Q_1, U_1)$ and $(Q_2, U_2)$ on the files $F_1$ and $F_2$ are obtained, due to $\text{Đ} \mid (\delta Q_1 - WH_1)$ and $\text{Đ} \mid (\delta Q_2 - WH_2)$, we see that

$$\text{Đ} \mid (\delta(Q_1 + Q_2) - W(H_1 + H_2)).$$

Let $Q' = Q_1 + Q_2$, $H' = H_1 + H_2$, then $\text{Đ} \mid (\delta Q' - WH')$. However, inferring $F'$ from $H'$ is intractable according to the properties of hash functions. In addition, finding a fit $U'$ from

$$U'^{T\text{đ}} \equiv ((\alpha^{-1}Q')^{Q'TS^{-1}}\bar{G}_1'^{Q'T})^{\text{đ}}\ (\%\ M)$$

is also intractable since $U'$ has $T\text{đ}$ values.

If many of the pair $(Q, U)$ are gathered, because $Q$ is random, $Q$ and $U$ interrelate through a transcendental logarithm, and the value of $U$ varies intensely between 1 and $M$, there is no polynomial function or statistic regularity among different $(Q, U)$'s, which indicates that they are unhelpful in solving TLP, but yet helpful in finding out the elements of the subgroup of order đ or $\text{đ}T$ as is pointed out in Section 6.2.

Thus, forging a signature through known signatures with a public key is of TLP or the $hash^{-1}(H)$.

## 6.5 Adaptive-chosen-message Attack Is Faced with Indistinguishability

In accordance with Section 3.4, $Q$ satisfies $\text{Đ} \mid (\delta Q - WH)$, namely $Q \equiv (\bar{a}\text{Đ} - WH)\delta^{-1}\ (\%\ \bar{M})$, where $\bar{a}$ is a random integer, and meets $(\text{đ}T) \nmid \bar{a}$.

The randomness of $\bar{a}$ leads $Q$ to be random while $U$ is interrelated with $Q$ in a transcendental logarithm, where $Q \equiv (RG_0)^S\delta\hbar\ (\%\ M)$, and $U \equiv (RW^{k-\delta})^Q\delta^{\bar{a}\text{Đ}r}\ (\%\ M)$.

Hence, for the identical file $F$, there will be many different signatures on it, which manifests that the signature $(Q, U)$ owns indistinguishability.

In terms of [38], the signature $(Q, U)$ on $F$ is secure against adaptive-chosen-message attack.

## 6.6 Chosen-signature Attack Is Faced with PRFP

It is well understood from the discriminant that

$$\bar{G}_1^{QUST}\beta^{HQ^{n-1} + H^n} \equiv (\alpha Q^{-1})^{QUT}\alpha^{Q^n}U^{UST}\ (\%\ M). \qquad (8)$$

If the values of $Q$ and $U$ are chosen in advance, an adversary may attempt to figure out $H$ and the corresponding file or message $F$.

Let $\bar{G}_1 = f(H) = \prod_{i=1}^{n} C_i^{b_i}\ \%\ M$, where $H = b_1 \ldots b_n = \sum_{i=1}^{n} b_i 2^{n-i}$, then (8) is an equation in $b_1, \ldots, b_n$.

Furthermore, let $g$ be a generator of $(\mathbb{Z}_M^*, \cdot)$, and through the Index-calculus method, work out $q_1, \ldots, q_n$, $v$, $u$, $w$ such that

$$g^{q_1} \equiv C_1 \ (\%\ M), \ \ldots, \ g^{q_n} \equiv C_n \ (\%\ M), \ g^v \equiv \beta \ (\%\ M),$$
$$g^u \equiv (\alpha Q^{-1})^{QUT} \alpha^{Q^n} U^{UST} \ (\%\ M), \ w \equiv QUST \ (\%\ \bar{M}).$$

Then, there is

$$(q_1 b_1 + \ldots + q_n b_n) w + v(Q^{n-1} \textstyle\sum_{i=1}^n b_i 2^{n-i} + (\sum_{i=1}^n b_i 2^{n-i})^n) \equiv u \ (\%\ \bar{M}),$$

which is of the multivariate polynomial root finding problem (Multivariate PRFP). Clearly, even though $H$ is found, it is infeasible to infer a fit $F$ from $H$.

On the other hand, if there exists the inverse function $H = f^{-1}(\bar{G}_1)$, namely $H$ in (8) is substituted with $\bar{G}_1$, evaluating $\bar{G}_1$ from (8) is the combination of RFP, DLP and PRFP.

## 7 Conclusion

REESSE1+ is only a prototypal cryptosystem which is used for explaining some concepts, ideas, and methods, so the space and time complexities of the five algorithms are not analyzed in the paper.

A REESSE1+ private key contains $2n + 5$ variables, but does not contain quadratic polynomials; thus REESSE1+ is a multivariate cryptosystem different from TTM and TTS.

In REESSE1+, not only the numerical calculation ability but also the logic judgement ability of a computer is utilized; thus the reversibility of the functions is relatively poor.

MPP which contains indeterminacy is a composite problem integrating IFP with DLP. ASPP is also a composite problem integrating IFP, DLP with ASSP, where ASSP can resist the $L^3$ lattice base reduction. TLP is a primitive problem which may be regarded as consisting of two variables. PRFP is also a primitive problem which contains both addition and multiplication — $ax^n + bx^{n-1} + cx + d \equiv 0 \ (\%\ \bar{M})$ with $n \ge 80$, $a \ne 0, 1$, $d \ne 0$, and $|b| + |c| \ne 0$ for example. So far, a generic subexponential algorithm for solving MPP, ASPP, TLP, or PRFP is not found. Due to indeterminacy, even as $\lg M \approx 80$, solving MPP, ASPP, TLP, or PRFP for the original answer is infeasible in subexponential time yet. Notice that $\lg M \approx 80$ indicates that the constraint $M > (max_{1 \le i \le n} A_i)^n$ is removed from the key generator, and REESSE1+ is only used for digital signature.

Some evidence given in the paper inclines people to believe that MPP, ASPP, and TLP are harder than DLP in the same prime field $\mathbb{GF}(M)$ each, and the evidence given in [47] inclines people to believe that PRFP is harder than RFP, which makes people be interested in it whether or not there exists a polynomial time algorithm for solving MPP, ASPP, TLP, or PRFP in the quantum computational model [21].

At present, the REESSE1+ cryptosystem is constructed in a prime field $\mathbb{Z}_M$, namely $\mathbb{GF}(M)$.

Suppose that $M$ is still a prime number. Then $\mathbb{Z}_M$ is a finite field with general addition and multiplication, and $\mathbb{Z}_M[x]$ is a Euclidean domain over $\mathbb{Z}_M$, namely a principle ideal domain and a uniquely factorial domain [22]. Additionally, we suppose that $P(x) \in \mathbb{Z}_M[x]$ is an irreducible polynomial of which the coefficient of the first term is the integer 1. Then $\mathbb{Z}_M[x] \,/\, P(x)$ constitutes a polynomial ring including a congruent Abelian group. Therefore, it is feasible to transplant REESSE1+ to the ring $\mathbb{Z}_M[x] \,/\, P(x)$ from the prime field $\mathbb{Z}_M$.

From the dialectical viewpoint, it is impossible that a public key cryptosystem possesses all the merits because some merits are possibly restrained by others. Along with the development of CPU techniques and quantum computations, what people are more concerned about are the securities of cryptosystems, but not the lengths of parameters.

Clearly, as viewed from utility, it should be researched further how to decrease the length of a REESSE1+ modulus and to increase the speed of a REESSE1+ decryption.

## Acknowledgment

The authors would like to thank the Academicians Jiren Cai, Zhongyi Zhou, Jianhua Zheng, Changxiang Shen, Zhengyao Wei, Andrew C. Yao, Binxing Fang, Guangnan Ni, and Xicheng Lu for their important guidance, advice, and suggestions.

The authors also would like to thank the Professors Dingyi Pei, Jie Wang, Ronald L. Rivest, Moti Yung, Dingzhu Du, Mulan Liu, Huanguo Zhang, Dengguo Feng, Yixian Yang, Maozhi Xu, Qibin Zhai, Hanliang Xu, Xuejia Lai, Yongfei Han, Kefei Chen, Yupu Hu, Rongquan Feng, Ping Luo, Dongdai Lin, Jianfeng Ma, Lei Hu, Lusheng Chen, Xiao Chen, Wenbao Han, Lequan Min, Bogang Lin, Xiulin Zheng, Hong Zhu, Renji Tao, Bingru Yang, Zhiying Wang, Quanyuan Wu, and Zhichang Qi for their important counsel, suggestions, and corrections.

# Authors

**Shenghui Su** received a bachelor degree in computer science from National University of Defense Technology, a master degree from Peking University, and a Ph.D. degree from University of Science and Technology Beijing. He was given the title of a senior programmer by Ministry of Electronic Industry China in 1994, and of a professor by Academic and Technologic Committee Beijing in 2004. He was long engaged in design of algorithms and development of software. Since 2000, he has been indulged in designing the REESSE / JUNA cryptosystems and digital signers. He is currently with college of computers, Beijing University of Technology. His research area covers computational complexity, cryptographic algorithms, and digital identities.

**Shuwang Lü** received a bachelor degree in electronics from University of Science and Technology China. He sits on the academic committee of SKLOIS. Since 1980, he has mainly been making researches on cipher algorithms and cipher chips. He presided over the projects supported by the national plan of high technology development, and of fundamental science researches, and was the chief designer of the SMS4 symmetric cryptosystem, an industrial standard. He is with the School of Graduate, Chinese Academy of Sciences, with University of Science and Technology China, and with USTB as a professor and a Ph.D. advisor. Nowadays, his research interests are still focused on cryptographic algorithms and knowledge security.

## Appendix A — Indeterminate Encryption

Below, a small example is given.

This example is only used to explain how to encrypt a plaintext and decrypt a ciphertext through the algorithms in Section 5.3.1, and thus the quantities $ħ$, $S$, $\alpha$, $\beta$ and the some constraints are not considered.

Let $n = 6$ and $Ƀ = 19$.

### 1) Generation of a key pair

① Select $đ = 21$, $Đ = 95$, and $T = 143$.

② Randomly generate a coprime sequence $\{A_i\} = \{17, 10, 13, 9, 19, 7\}$.

③ Find the prime $M = 174594421$ such that $M > 19^6$ and $(đĐT) \mid (\bar{M} = 174594420)$.

④ Pick $W = 155629$ and $\delta = 3761$ making $\gcd(\delta, \bar{M}) = 1$ and $\|\delta\| = đĐT$.

⑤ Randomly produce pairwise distinct $\{\ell(i)\} = \{7, 15, 5, 11, 13, 9\}$.

⑥ Compute $\{C_i\} = \{116331875, 87811986, 61498911, 6213388, 8089113, 9766243\}$ by $C_i \leftarrow (A_i W^{\ell(i)})^{\delta}$ % $M$.

Regard ($\{C_i\}$, $M$) as a public key, and ($\{A_i\}$, $W$, $\delta$) as a private key. Discard the quantities $đ$, $Đ$, $\{\ell(i)\}$ which cannot be divulged.

### 2) Encryption

Assume that ($\{C_i\}$, $M$) is a public key. Let $b_1 \ldots b_6 = 100110$ be a plaintext.

① Obtain $ƀ_1 \ldots ƀ_6 = 100310$, $k = 1$.

② Randomly generate $s_1 \ldots s_6 = 010001$.

③ Modify $s_5$ and $ƀ_5$ by $s_5 \leftarrow 1$ and $ƀ_5 \leftarrow ƀ_5 + k = 2$.

④ Compute $\bar{G} \equiv (C_{1\times 0+(1-1+1)\neg 0})^1 (C_{2\times 1+(2-0+1)\neg 1})^0 (C_{3\times 0+(3-0+1)\neg 0})^0 (C_{4\times 0+(4-3+1)\neg 0})^3 (C_{5\times 1+(5-2+1)\neg 1})^2 (C_{6\times 1+(6-0+1)\neg 1})^0 \equiv C_1^{\,1} C_2^{\,3} C_5^{\,2} \equiv 116331875 \times 156599315 \times 124996494 \equiv 75924783$ (% $M$) by $\bar{G} \equiv \prod_{i=1}^{n} (C_{i s_i + (i - ƀ_i + 1)\neg s_i})^{ƀ_i}$ (% $M$) with $C_{n+1} = 1$.

So, the ciphertext $\bar{G} = 75924783$ is obtained. Notice that because $s_1 \ldots s_6$ is randomly generated, on inputting the identical plaintext 100110 many times, we will obtain many distinct ciphertexts.

### 3) Decryption

Assume that ($\{17, 10, 13, 9, 19, 7\}$, 155629, 3761) is a related private key.

Let $\bar{G} = 75924783$ be a ciphertext.

① Compute $\bar{G} \equiv \bar{G}^{\delta^{-1}} \equiv 75924783^{3761^{-1}} \equiv 75924783^{4781501} \equiv 165482231$ (% $M$).
Compute $W^{-2} \equiv (W^{-1})^2 \equiv (1171225)^2 \equiv 154229249$ (% $M$).

Outer loop 1:

② Compute $\bar{G} \equiv 165482231 \times 154229249 \equiv 144398410$ (% $M$) by $\bar{G} \leftarrow \bar{G}W^{-2}$ % $M$.

③ Set $b_1 \ldots b_6 = 0 \ldots 0$, $ë = 0$, $j = 0$, $k = 0$, $G = \bar{G} = 144398410$, $i = 1$.

Inner loop 1:

④ Due to $(A_1^{\,ë+1} = 17^1) \nmid (G = 144398410)$, the next.

⑤ Due to $ë = 0$,
let $k = 0 + 1 = 1$, $i = 1 + 1 = 2$.

⑥ Due to $i = 2 \le 6$ and $G \ne 1$, goto ④.

Inner loop 2:

④ Due to $(A_2^{\,ë+1} = 10^1) \mid (G = 144398410)$, $ë = 0 + 1 = 1$, goto ④.

④ Due to $(A_2^{\,ë+1} = 10^2) \nmid (G = 144398410)$, the next.

⑤ Due to $ë = 1 \ne 0$,
compute $G = 144398410 / A_2^{\,1} = 14439841$;
owing to $k = 1 > 0$, let $b_2 = 1$;
owing to $k = 1 \ne 0$, let $i = 2 + 1 = 3$;
owing $k + 1 = 2 > ë = 1$, let $i = 6 + 1 = 7$;
set $ë = 0$, $k = 0$.

⑥ Due to $i = 7 > 6$, the next.

⑦ Due to $G = 14439841 \ne 1$, goto ②.

Outer loop 2:

⋮

⋮

Outer loop 39:

② Compute $\bar{G} \equiv 131367179 \times 154229249 \equiv 6137000$ (% $M$) by $\bar{G} \leftarrow \bar{G}W^{-2}$ % $M$.

③ Set $b_1 \ldots b_6 = 0 \ldots 0$, $ë = 0$, $j = 0$, $k = 0$, $G = \bar{G} = 6137000$, $i = 1$.

Inner loop 1:

④ Due to $(A_1^{\,ë+1} = 17^1) \mid (G = 6137000)$, $ë = 0 + 1 = 1$, goto ④.

④ Due to $(A_1^{\,ë+1} = 17^2) \nmid (G = 6137000)$, the next.

⑤ Due to $ë = 1 \ne 0$,
compute $G = 6137000 / A_1^{\,1} = 361000$;
owing to $k = 0$ and $i + ë - 1 < 6$, let $b_{i+ë-1} = b_1 = 1$;
owing to $k = 0$, let $i = 1 + ë = 1 + 1 = 2$;
owing to $k + 1 = 1 = ë$, the next;
set $ë = 0$, $k = 0$.

⑥ Due to $i = 2 \le 6$ and $G \ne 1$, goto ④.

Inner loop 2:

④ Due to $(A_2^{\,ë+1} = 10^1) \mid (G = 361000)$, $ë = 0 + 1 = 1$, goto ④.

④ Due to $(A_2^{\,ë+1} = 10^2) \mid (G = 361000)$, $ë = 1 + 1 = 2$, goto ④.

④ Due to $(A_2^{\,ë+1} = 10^3) \mid (G = 361000)$, $ë = 2 + 1 = 3$, goto ④.

④ Due to $(A_2^{\,ë+1} = 10^4) \nmid (G = 361000)$, the next.

⑤ Due to $ë = 3 \ne 0$,
compute $G = 361000 / A_2^{\,3} = 361$;
owing to $k = 0$ and $i + ë - 1 < 6$, let $b_{i+ë-1} = b_4 = 1$;
owing to $k = 0$, let $i = 2 + ë = 2 + 3 = 5$;
owing to $k + 1 = 1 < ë = 3$, the next;
set $ë = 0$, $k = 0$.

⑥ Due to $i = 5 \le 6$ and $G \ne 1$, goto ④.

Inner loop 3:

④ Due to $(A_5^{\,ë+1} = 19^1) \mid (G = 361)$, $ë = 0 + 1 = 1$, goto ④.

④ Due to $(A_5^{\,ë+1} = 19^2) \mid (G = 361)$, $ë = 1 + 1 = 2$, goto ④.

④ Due to $(A_5^{\,ë+1} = 19^3) \nmid (G = 361)$, the next.

⑤ Due to $ë = 2 \ne 0$,
compute $G = 361 / A_5^{\,2} = 1$;
owing to $i + ë - 1 = 5 + 2 - 1 = 6$, let $b_i = b_5 = 1$;
owing to $k = 0$, let $i = 5 + 2 = 7$;
owing to $k + 1 = 1 < ë = 2$, the next;

⑥ Due to $i = 7 > 6$ or $G = 1$, the next.

⑦ Due $G = 1$, end.

In this way, we recover the original plaintext $b_1 \ldots b_6$ = 100110.

## Appendix B — Computation of Density of a Knapsack from ASPP

### 1) Wrong Computation of Density in Section 5.2

It is known from Section 3.2 that a ciphertext is an ASPP $\bar{G} \equiv \prod_{i=1}^{n} C_i^{\underset{.}{b}_i}$ (% $M$).

Let $C_1 \equiv g^{u_1}$, …, $C_n \equiv g^{u_n}$, $\bar{G} \equiv g^{v}$ (% $M$), where $g$ is a generator of $(\mathbb{Z}_M^*, \cdot)$ randomly selected.

Then, seeking $\underset{.}{b}_1 \ldots \underset{.}{b}_n$ from $\bar{G}$ is equivalent to solving the congruence

$$u_1 \underset{.}{b}_1 + \ldots + u_n \underset{.}{b}_n \equiv v \ (\% \ \bar{\boldsymbol{M}}), \qquad (1)$$

where $\{u_1, \ldots, u_n\}$ is called a compact sequence (knapsack) due to $\underset{.}{b}_i \in [0, n/2+1]$. Seeking $\underset{.}{b}_1 \ldots \underset{.}{b}_n$ from (1) is called the anomalous subset sum problem (ASSP).

Note that ① we stipulate that $b_1 \ldots b_n \neq 0$ contains at most $n/2$ 0-bits; ② if $g$ is different, $\{u_1, \ldots, u_n\}$ will be different for the same $\{C_1, \ldots, C_n\}$, and thus $\{u_1, \ldots, u_n\}$ has randomicity.

When (1) will be reduced through the LLL lattice basis reduction algorithm, it should be converted into a non-modular form:

$$\underset{.}{b}_1 u_1 + \ldots + \underset{.}{b}_n u_n \equiv v + k\bar{\boldsymbol{M}}, \qquad (2)$$

where $k \in [0, n]$ is an integer. To seek the original solution to (2), $k$ must traverse from 0 to $n$.

Let $D$ be the density of the compact sequence $\{u_1, \ldots, u_n\}$. We easily see that in Section 5.2, the formula $D \approx n^2 / \lceil \lg M \rceil$ is wrong, which is first pointed out by Xiangdong Fei (an associate professor from Nanjing University of Technology).

### 2) Right Computation of Density

Considering the structure of a lattice basis from (1) and the bit-length of a bit shadow $\underset{.}{b}_i$, the right computation of density of an ASSP knapsack should be

$$D = \sum_{i=1}^{n} \lceil \lg(n/2+1) \rceil / \lceil \lg M \rceil = n \lceil \lg(n/2+1) \rceil / \lceil \lg M \rceil$$

with $\underset{.}{b}_i \in [0, n/2+1]$

or

$$D = \sum_{i=1}^{n} \lceil \lg n \rceil / \lceil \lg M \rceil = n \lceil \lg n \rceil / \lceil \lg M \rceil \text{ with } \underset{.}{b}_i \in [0, n]$$

Assume that $\underset{.}{b}_i \in [0, n]$. Concretely speaking,

for $n = 80$ with $\lceil \lg M \rceil = 696$, $D = 80 \times 7 / 696 \approx 0.8046 < 1$;

for $n = 96$ with $\lceil \lg M \rceil = 864$, $D = 96 \times 7 / 864 \approx 0.7778 < 1$;

for $n = 112$ with $\lceil \lg M \rceil = 1030$, $D = 112 \times 7 / 1030 \approx 0.7612 < 1$;

for $n = 128$ with $\lceil \lg M \rceil = 1216$, $D = 128 \times 8 / 1216 \approx 0.8421 < 1$.

These densities mean that the original solution to (1) may possibly be found through LLL lattice basis reduction (not certainly, and even with very low probability) because $D < 1$ only assure that the shortest vector is unique, but it cannot assure that the vector of the original solution to an ASSP is just the shortest vector in the reduced basis.

### 3) Bit Shadows Enhance Resistance of a Low Density ASSP Knapsack to Attacks

The LLL algorithm is to reduce a lattice basis $\langle 1, 0, \ldots, 0, \tilde{\boldsymbol{N}} u_1 \rangle$, $\langle 0, 1, \ldots, 0, \tilde{\boldsymbol{N}} u_2 \rangle$, …, $\langle 0, 0, \ldots, 1, \tilde{\boldsymbol{N}} u_n \rangle$, $\langle 1/2, 1/2, \ldots, 1/2, \tilde{\boldsymbol{N}}(v + k\bar{\boldsymbol{M}}) \rangle$, where $\tilde{\boldsymbol{N}} > 1/2(n)^{1/2}$. No matter whether $(v + k\bar{\boldsymbol{M}})$ is a classical subset sum or an anomalous subset sum, and whether the density is less than 1 or greater than 1, the LLL algorithm runs by its inherent rules. Lastly, the $n + 1$ vectors $\langle \hat{e}_1, \ldots, \hat{e}_n, \hat{e}_{n+1} \rangle$'s which occur in the reduced basis are the first $n + 1$ approximately shortest vectors, including the shortest vector, of which quite some satisfy $\hat{e}_1 u_1 + \ldots + \hat{e}_n u_n \equiv v + k\bar{\boldsymbol{M}}$ (omitting the term $\hat{e}_{n+1} = 0$). If $D < 1$, the shortest vector is unique; and if $(v + k\bar{\boldsymbol{M}})$ is a classical subset sum, the shortest vector is just the original solution.

We know from the above discussion that it has two necessary conditions to solve a SSP or ASSP through LLL lattice basis reduction: ① the vector of the original solution is the shortest; ② the shortest vector in the lattice is unique. $D < 1$ assures that the shortest vector is unique; and a classical subset sum assures that the vector of the original solution is just the shortest vector.

Return to (1). Even though the density of an ASSP knapsack is less than 1, the **original** solution is not necessarily found since $\underset{.}{b}_i \in [0, n/2+1]$ is a bit shadow (indicates that the **original** solution does not necessarily occur in the reduced basis), and there likely exist many solutions in the lattice.

For example, let $n = 4$ (short but without loss of generality), $M = 263$, $\{u_1, \ldots, u_4\} = \{48, 71, 257, 4\}$, and $v = 261$ ($u_i$ and $v$ are obtained through the discrete logarithms of $C_i$ and $\bar{G}$). Here, the density of $\{u_1, \ldots, u_4\}$ is $D = n \lceil \lg(n/2+1) \rceil / \lceil \lg M \rceil = 4 \times 2 / 9 = 0.8889 < 1$.

Assume that a plaintext $b_1 \ldots b_4$ = 1100, and its related bit shadow string $\underset{.}{b}_1 \ldots \underset{.}{b}_4$ = 1300 (thus $1 \times 48 + 3 \times 71 = 261$).

However, according to LLL lattice basis reduction, sought solution will be 0011 (notice, not the original) because there is $1 \times 257 + 1 \times 4 = 261$, and $\langle 0, 0, 1, 1 \rangle \mid (1^2 + 1^2)^{1/2}$ is the shortest vector in the lattice while $\langle 1, 3, 0, 0 \rangle \mid (1^2 + 3^2)^{1/2}$ is not the shortest, and even it will not occur in the reduced basis consisting of 5 vectors.

### 4) Density of an Optimized ASSP Knapsack

The modulus of prototypal REESSE1+ is relatively large, so in practice, it needs to be optimized (optimized REESSE1+ is called JUNA).

Return to Section 6. When $n$ = 80, 96, 112, 128, correspondingly there is $\lceil \lg M \rceil$ = 384, 464, 544, 640. The density of an optimized ASSP knapsack is $D = (3n/2) \lceil \lg(n/4+1) \rceil / \lceil \lg M \rceil$ with $\underset{.}{B}_i \in [0, n/4+1]$ (on the assumption that $B_1 \ldots B_{n/2}$ contains at most $n/4$ 00-pairs).

Concretely speaking,

for $n = 80$ with $\lceil \lg M \rceil = 384$, $D = 120 \times 5 / 384 \approx 1.5625 > 1$;

for $n = 96$ with $\lceil \lg M \rceil = 464$, $D = 144 \times 5 / 464 \approx 1.5517 > 1$;

for $n = 112$ with $\lceil \lg M \rceil = 544$, $D = 168 \times 5 / 544 \approx 1.5441 > 1$;

for $n = 128$ with $\lceil \lg M \rceil = 640$, $D = 192 \times 6 / 640 \approx 1.8000 > 1$.

Under the circumstances, owing to $D > 1$ (indicates there will exist many solutions to the ASSP, and even the shortest vector is also nonunique), it is impossible to find the original plaintext $b_1 \ldots b_n$ through LLL lattice basis reduction. Therefore, at present there exists no subexponential time solution to the ASPP used in the optimized encryption scheme.

It should be noted that even if $80 \leq n \leq 128$, the LLL

lattice basis reduction algorithm right cannot find the original solution to SSP when $D > 1$, which is proven by our experiments.

## Appendix C — Offering a Reward

It may be regarded as a type of proof by experiment.

Here, $n \ge 80$ is the length of a binary string $b_1 \ldots b_n \neq 0$ of which the bit-pair string is $B_1 \ldots B_{n/2}$ containing at most $n/4$ 00-pairs, the sign % denotes 'modulo', $\bar{M}$ means '$M-1$' with $M$ prime, and $\lg x$ means the logarithm of $x$ to the base 2.

The analysis in the paper shows that any effectual attack on REESSE1+ will be reduced to the solution of four intractabilities: a multivariate permutation problem (MPP), an anomalous subset product problem (ASPP), a transcendental logarithm problem (TLP), and a polynomial root finding problem (PRFP) so far.

It is well known that it is infeasible in subexponential time to find a large root to the PRFP $ax^n + bx^{n-1} + cx + d \equiv 0$ (% $M$) with $a \notin \{0, 1\}$, $|b| + |c| \neq 0$, $d \neq 0$, and $n$, $M$ large enough.

Let $n$=80, 96, 112, 128 with $\lceil \lg M \rceil$=384, 464, 544, 640 for the optimized REESSE1+ encryption scheme or with $\lceil \lg M \rceil$=80, 96, 112, 128 for the lightweight REESSE1+ signing scheme.

Assume that $(\{C_1, \ldots, C_{3n/2}\}, M)$ is a public key, and $(\{A_1, \ldots, A_{3n/2}\}, \{\ell(1), \ldots, \ell(3n/2)\}, W, \delta, M)$ with $W, \delta \in (1, \bar{M})$, $A_i \in \{2, 3, \ldots, 1201\}$, and $\ell(i) \in \{+/-5, +/-7, \ldots, +/-(2(3n/2) + 3)\}$ is a private key, where the sign +/– means that the plus sign + or minus sign – is selected, and unknown to the masses.

The authors promise solemnly that

① anyone who can extract the original private key definitely from the MPP

$$C_i \equiv (A_i W^{\ell(i)})^{\delta} \ (\%\ M) \text{ for } i = 1, \ldots, 3n/2$$

in DLP subexponential time will be awarded \$100000 when $n$ = 80, 96, 112, 128 with $\lceil \lg M \rceil$ = 384, 464, 544, 640, or \$10000 with $\lceil \lg M \rceil$ = 80, 96, 112, 128;

② anyone who can recover the original plaintext $b_1 \ldots b_n$ definitely from the ASPP

$$\bar{G} \equiv \prod_{i=1}^{n/2} (C_{3(i-1)+\underline{B}_i})^{\underline{B}_i} \ (\%\ M)$$

with $C_0 = 1$ and $\underline{B}_i$ a bit-pair shadow in DLP subexponential time will be awarded \$100000 when $n$ = 80, 96, 112, 128 with $\lceil \lg M \rceil$ = 384, 464, 544, 640, or \$10000 with $\lceil \lg M \rceil$ = 80, 96, 112, 128, where $\underline{B}_i = 0$ if $B_i$ = 00, = 1 + the number of successive 00-pairs before $B_i$ if $B_i \neq$ 00, or = 1 + the number of successive 00-pairs before $B_i$ + the number of successive 00-pairs after the rightmost non-00-pair if $B_i$ is the leftmost non-00-pair as $b_1 \ldots b_{12}$ = 010000110100 = $B_1 \ldots B_6$ = 01 00 00 11 01 00 with $\underline{B}_1 \ldots \underline{B}_6$ = 2 0 0 3 1 0.

③ anyone who can find the original large answer $x \in (1, \bar{M})$ definitely to the TLP

$$y \equiv (gx)^x \ (\%\ M)$$

with known $g, y \in (1, \bar{M})$ in DLP subexponential time will be awarded \$100000 when $n$ = 80, 96, 112, 128 with $\lceil \lg M \rceil$ = 384, 464, 544, 640, or \$10000 with $\lceil \lg M \rceil$ = 80, 96, 112, 128.

Of course, any solution must be described with a formal process, and can be verified with our examples. The time of solving a problem should be relevant to arithmetic steps, but irrelevant to CPU speeds.

The DLP subexponential time means the running time of an algorithm for solving the DLP in the prime field $\mathbb{GF}(M)$ through Index-calculus method at present, namely $L_M[1/3, 1.923]$.

Note that the TLP is written as $y \equiv (gx)^x$ (% $M$) instead of $y \equiv x^x$ (% $M$) due to the asymptotic property of $M$, and in the paper, some pieces of evidence found incline people to believe that the subset product problem (SPP) $\bar{G}_1 \equiv \prod_{i=1}^{n} C_i^{b_i}$ (% $M$) is harder than the DLP asymptotically, but due to $\lceil \lg M \rceil \le 640$ and the density of a related knapsack being low, SPP can almost be solved in DLP subexponential time.